\documentclass[a4paper,oneside,10pt]{amsart}

\usepackage[T1]{fontenc}
\usepackage{amsmath,amssymb,epsf}
\usepackage{amsfonts,amsbsy,amscd,stmaryrd}
\usepackage{latexsym,amsopn}
\usepackage{amsthm}
\usepackage{esint}
\usepackage{mathrsfs}

\usepackage{graphicx}
\usepackage{color}
\definecolor{Blue}{rgb}{0.,0.,1.}
\definecolor{Red}{rgb}{1.,0.,0.}

\newcounter{smallarabics}
\newenvironment{arabicenumerate}
{\begin{list}{{\normalfont\textrm{(\arabic{smallarabics})}}}
  {\usecounter{smallarabics}\setlength{\itemindent}{0cm}
   \setlength{\leftmargin}{5ex}\setlength{\labelwidth}{4ex}
   \setlength{\topsep}{0.75\parsep}\setlength{\partopsep}{0ex}
   \setlength{\itemsep}{0ex}}}
{\end{list}}

\newcounter{smallroman}

\newenvironment{notations}
{\begin{list}{{\normalfont\textrm{-}}}
  {\setlength{\itemindent}{0cm}
   \setlength{\leftmargin}{2ex}\setlength{\labelwidth}{4ex}
   \setlength{\topsep}{0.75\parsep}\setlength{\partopsep}{1ex}
   \setlength{\itemsep}{1ex}}
}
{\end{list}}

\let\origmaketitle\maketitle
\def\maketitle{
  \begingroup
  \def\uppercasenonmath##1{} 
  \let\MakeUppercase\relax 
	\origmaketitle
  \endgroup
}

\newcommand{\ben}{\begin{arabicenumerate}}  
\newcommand{\een}{\end{arabicenumerate}}

\def\init{\setcounter{equation}{0}}

\newtheorem{theorem}{Theorem}[section]

\newtheorem{proposition}[theorem]{Proposition}
\newtheorem{lemma}[theorem]{Lemma}
\newtheorem{definition}[theorem]{Definition}

\newtheorem{remark}[theorem]{Remark}
\newtheorem{example}[theorem]{Example}
\newcommand{\beq}{\begin{equation}}
\newcommand{\eeq}{\end{equation}}

\newcommand{\bex}{\begin{example}}
\newcommand{\eex}{\end{example}}
\def\bel{\begin{lemma}}
\def\eel{\end{lemma}}
\def\bet{\begin{theorem}}
\def\eet{\end{theorem}}
\def\bed{\begin{definition}}
\def\eed{\end{definition}}
\def\ber{\begin{remark}}
\def\eer{\end{remark}}

\def\rr{{\mathbb R}}

\def\cc{{\mathbb C}}
\def\nn{{\mathbb N}}

\def\slim{{\rm s-}\lim}

\def\bar{\overline}

\def\r{{\rm r}}
\def\cinf{C^\infty}
\def\c0inf{C_0^\infty}

\def\proof{
\noindent{\bf Proof.}\ \ }

\usepackage{calrsfs}
\DeclareMathAlphabet{\pazocal}{OMS}{zplm}{m}{n}
\DeclareMathAlphabet{\mathsfsl}{OMS}{cmss}{m}{n}

\DeclareSymbolFont{altletters}  {OML}{zplm}{m}{it}
\DeclareMathSymbol{\altdelta}{\mathalpha}{altletters}{"0E}
\DeclareMathSymbol{\alteta}{\mathalpha}{altletters}{"11}

\def\cY{{\pazocal Y}}

\def\cD{{\mathcal D}}
\def\cU{{\mathcal U}}

\def\cN{{\pazocal N}}

\def\cW{{\pazocal W}}

\def\wf{{\rm WF}}

\def\free{{{\rm free}}}

\def\i{{\rm i}}

\DeclareMathOperator{\Dom}{Dom}

\def\vac{{\rm vac}}

\def\qed{$\Box$\medskip}
\newcommand{\qeds}{\qed}

\DeclareMathOperator{\ind}{ind}

\DeclareMathOperator{\Ker}{Ker}
\DeclareMathOperator{\Ran}{Ran}
\DeclareMathOperator{\coKer}{coKer}

\def \p{ \partial}
\def\12{\frac{1}{2}}
\def\14{\frac{1}{4}}

\def\supp{{\rm supp}}
\def\e{{\rm e}}
\newcommand{\one}{\boldsymbol{1}}
\def\cH{{\pazocal H}}

\def\coinf{C_{\rm c}^\infty}

\def\c{{\pazocal }}

\def\cF{{\pazocal F}}

\def\cX{{\pazocal X}}

\def\cK{{\pazocal K}}

\def\12{\frac{1}{2}}

\def\supp{{\rm supp}}

\def\e{{\rm e}}

\def\Diff{{\rm Diff}}
\def\rx{{\rm x}}
\def\bx{{\rm x}}

\def\bep{\begin{proposition}}
\def\eep{\end{proposition}}

\def\Op{{\rm Op}^{\rm w}}

\newcommand{\mat}[4]{\begin{pmatrix}#1 &#2  \\ #3 &#4 \end{pmatrix}}

\def\CARal{{\rm C\hskip 0.25 em \hbox{\raise 1.72 ex 
\hbox{$\scriptscriptstyle\rm al$}\kern -0.57 em A}R}}

\def\otimesal{\mathop{\hbox{\raise 1.5 ex
  \hbox{$\scriptscriptstyle\rm al$}
\kern -0.92 em \hbox{$\otimes$}}}}
\def\oplusal{\mathop{\hbox{\raise 1.5 ex
  \hbox{$\scriptscriptstyle\rm al$}
\kern -0.92 em \hbox{$\oplus$}}}}
\def\Gammal{\hbox{\raise 1.68 ex 
\hbox{$\scriptscriptstyle\rm al$}\kern -0.50 em $\Gamma$}}
\def\Bal{\hbox{\raise 1.68 ex 
\hbox{$\scriptscriptstyle\rm  al$}\kern -0.50 em $B$}}
\def\CARal{{\rm C\hskip 0.25 em \hbox{\raise 1.72 ex 
\hbox{$\scriptscriptstyle\rm al$}\kern -0.57 em A}R}}

\def\cE{\pazocal{E}}

\DeclareMathAlphabet{\mathpzc}{OT1}{pzc}{m}{it}
\newcommand{\bra}{\langle} 
\newcommand{\ket}{\rangle}

\DeclareSymbolFont{boldoperators}{OT1}{cmr}{bx}{n}
\SetSymbolFont{boldoperators}{bold}{OT1}{cmr}{bx}{n}

\makeatletter
\newcommand*{\defeq}{\mathrel{\rlap{%
                     \raisebox{0.3ex}{$\m@th\cdot$}}%
                     \raisebox{-0.3ex}{$\m@th\cdot$}}%
                     =}
                     
\makeatother
\makeatletter
\newcommand*{\eqdef}{=\mathrel{\rlap{%
                     \raisebox{0.3ex}{$\m@th\cdot$}}%
                     \raisebox{-0.3ex}{$\m@th\cdot$}}%
                     }
\makeatother
\DeclareMathAlphabet{\mathpzc}{OT1}{pzc}{m}{it}

\def\Op{{\rm Op}}
\def\WF{{\rm WF}}

\def\altch{\skew3\hat{\rm\textit{c}}}

\def\altg{{\rm\textit{g}}}
\def\alth{{\rm\textit{h}}}
\def\altV{{\rm\textit{V}}}

\def\altm{{\rm\textit{m}}}
\def\altVh{\skew5\hat{\rm\textit{V}}}

\newcommand{\bea}{\begin{aligned}}
\newcommand{\beal}{\begin{array}{l}}
\newcommand{\eeal}{\end{array}}
\newcommand{\eea}{\end{aligned}}

\def\cf{C^\infty}

\def\td{{\rm td}}
\def\std{{\rm std}}

\def\Hstd{{\rm std}}
\def\aM{{\rm aM}}
\def\scc{{\rm sd}}

\def\nt{{\rm nt}}

\def\adg{{\rm ad}}
\def\dg{{\rm d}}

\def\rf{{\rm ref}}

\def\sca{{\rm out/in}}

\def\inout{{\rm in/out}}
\def\inn{{\rm in}}
\def\out{{\rm out}}
\def\F{{\rm F}}
\def\aF{{\rm \overline{F}}}

\def\spexi{{k}}

\newcommand{\traa}[1]{\mskip-6mu\upharpoonright_{#1}}
\def\pe{\overline{\p}}

\def\zero{{\mskip-4mu{\rm\textit{o}}}}
\def\diag{{\rm diag}}
\def\ry{{\rm y}}

 \def\gdia{G^{\dg}}
\def\outin{{\rm out/in}}
\def\varo{\varrho}
\def\varT{t_{+}}
\def\sd{{\rm sd}}
\def\sobo{{m}}
\def\AH{H}

\begin{document}
\title[The massive Feynman propagator on asymptotically  Minkowski spacetimes]{\Large The massive Feynman propagator \\ on asymptotically  Minkowski spacetimes}
\author{}
\address{Universit\'e Paris-Sud XI, D\'epartement de Math\'ematiques, 91405 Orsay Cedex, France}
\email{christian.gerard@math.u-psud.fr}
\author{\normalsize Christian \textsc{G\'erard} \& Micha{\l} \textsc{Wrochna}}
\address{Universit\'e Grenoble Alpes, Institut Fourier, UMR 5582 CNRS, CS 40700, 38058 Grenoble \textsc{Cedex} 09, France}
\email{michal.wrochna@ujf-grenoble.fr}
\begin{abstract}We consider the massive Klein-Gordon equation on asymptotically Minkowski spacetimes, in the sense that the manifold is $\rr^{1+d}$ and the metric approaches that of Minkowski space at infinity in a short-range way (jointly in time and space variables). In this setup we define Feynman and anti-Feynman scattering data and prove the Fredholm property of the Klein-Gordon operator with the associated Atiyah-Patodi-Singer type boundary conditions at infinite times. We then construct a parametrix (with compact remainder terms) for the Fredholm problem and prove that it is also a Feynman parametrix in the sense of Duistermaat and H\"ormander.
\end{abstract}

\maketitle

\section{Introduction \& summary}

\subsection{Introduction}

In the analysis of the free Klein-Gordon equation
\beq\label{eq:KG1}
(\p_{t}^{2}- \Delta_{\rx}+ \altm^{2})u(t,\rx)=0 \ \mbox{ on } \rr_t\times\rr^{d}_\rx,
\eeq
an essential elementary fact is that the Klein-Gordon operator $P_\free=\p_{t}^{2}- \Delta_{\rx}+ \altm^{2}$ possesses four distinguished inverses, namely the operators that multiply the Fourier transform of distributions by multiples of respectively
\[
\frac{1}{(\tau\pm \i 0 )^2 - (k^2+\altm^2)}, \ \  \frac{1}{\tau^2 - (k^2+\altm^2)\pm\i0},
\]
and which are called the retarded/advanced, resp. Feynman/anti-Feynman propagator (we use the notation $\xi=(\tau,k)$ for the covariables corresponding to $x=(t,\rx)\in\rr^{1+d}$).

The retarded and advanced propagators are intimately linked to the Cauchy problem for \eqref{eq:KG1}, and can also be equivalently defined as the unique operators that solve the retarded/advanced problem
\beq\label{eq:KG2}
P_\free u =f, \ \ \supp\, u\subset(\supp f)\pm C_+,
\eeq
where $C_+$ is the forward lightcone.
The well-posedness of the advanced/retarded problem \eqref{eq:KG2} is a fact that generalizes to setups such as the Klein-Gordon operator
\[
P\defeq -\Box_\altg + \altV
\]
on a globally hyperbolic\footnote{Let us recall that $(M,g)$ is globally hyperbolic if it admits a Cauchy surface, i.e., a smooth hypersurface that is intersected by every inextensible, non-spacelike (i.e. causal) curve exactly once.} spacetime $(M,\altg)$ and with potential $V\in\cf(M;\rr)$ (see e.g. \cite{BGP}), and in consequence, this provides a natural and unambiguous definition of advanced and retarded propagators in a broad range of situations. 

On the other hand, it is hardly obvious how the Feynman and anti-Feynman propagators generalize. This is a problem central to Quantum Field Theory on curved spacetimes in view of the role played by the Feynman propagator in interacting QFTs on Minkowski space. 

A  groundbreaking advance in that respect was provided by the work of Duistermaat and H\"ormander \cite{DH}, who proved the existence and uniqueness modulo smooth terms of \emph{Feynman} and \emph{anti-Feynman parametrices}, i.e. inverses of $P$ modulo smooth (but not necessarily compact) terms, distinguished by a specific structure of the singularities of the Schwartz kernel, as described by its \emph{wave front set}.  For later reference, let us recall the precise formulation:

\begin{definition}\label{def:Fp} We say that $G_\F$ is a Feynman parametrix if the operators $\one- G_\F P$ and $\one-P G_\F$ have smooth Schwartz kernel and
\beq\label{eq:fewf}
\WF'(G_{\F})= (\diag_{T^*M})\cup\textstyle\bigcup_{s\leq 0}(\Phi_s(\diag_{T^*M})\cap \pi^{-1}\cN),
\eeq
where $\Phi_s$ is the bicharacteristic flow acting on the left component of $\diag_{T^*M}$ (i.e., the diagonal in $(T^*M\times T^*M)\setminus\zero$), and $\pi:\cN\times\cN\to\cN$ is the projection to the left component.
\end{definition}

Above, $\wf'(G_{\F})$ stands for the primed wave front set of $G_{\F}$, i.e. it is the image of the wave front set of the Schwartz kernel of $G_{\F}$ by the map $(x,\xi,x',\xi')\mapsto (x,\xi,x',-\xi')$. We refer to \cite{hoermander} for the definition and the basic properties of the wave front set of a distribution. We recall that the \emph{bicharacteristic flow} $\Phi_t$ is the Hamilton flow of $p(x,\xi)=\xi\cdot \altg^{-1}(x)\xi$ restricted to the \emph{characteristic set} $\cN=p^{-1}(\{0\})$ (understood as a subset of $T^*M\setminus\zero$, where $\,\zero$ is the zero section of the cotangent bundle), see \cite{hoermander}.

Although not directly applicable in QFT, where actual inverses and bi-solutions are needed rather than parametrices, the results of Duistermaat and H\"ormander were successfully adapted by Radzikowski in the study of so-called {\em Hadamard states} and of their two-point functions \cite{radzikowski}. This has triggered important developments, culminating in rigorous treatments of perturbative interacting QFT on curved spacetimes \cite{BF00,HW1,HW2,dang}, where the role of the Feynman propagator is played by a `time-ordered expression' obtained from Hadamard two-point functions,  see e.g. \cite{HW,KM} for recent reviews. Without going into details, let us point out that this gives a notion of Feynman propagators that satisfy \eqref{eq:fewf} and which are inverses of $P$ (in the sense that when composed with $P$ acting on test functions, they give the identity), and are constructed using methods that are \emph{local} or \emph{asymptotic} in time\footnote{A notable exception  is a global construction that is shown to work for static and cosmological space-times in \cite{brumfredenhagen} and which is based on an improvement of a simple argument from spectral theory discussed e.g. in \cite{FV} (cf. \cite{FMR} for the analogue in the case of the Dirac equation); its outcome is however highly non-unique.}. However, these are non-unique even  if possible asymptotic symmetries of $(M,\altg)$ are implemented. For instance, in a space-time that is asymptotic to Minkowski space as $t\to+\infty$ and $t\to-\infty$, there are at least two such Feynman propagators, one constructed from scattering data at $t=+\infty$ and the other one from $t=-\infty$ data \cite{inout}.

A dramatically different perspective was proposed recently by Gell-Redman, Haber and Vasy \cite{GHV} (basing on earlier developments including \cite{BVW,HV,semilinear,resolvent,kerrds}, cf. \cite{VW} for the proof of \eqref{eq:fewf} in that setting), who showed in the case of asymptotically Minkowski spacetimes (and assuming $\altV=0)$ that a Feynman parametrix can be obtained as the solution of a \emph{global} problem of the form $Pu=f$, with $u$ and $f$ being in carefully chosen Hilbert spaces of distributions, and $u$ being the unknown. This problem can be solved modulo finite dimensional, smooth terms (these can even be proved to vanish under extra assumptions), and therefore this yields a Feynman parametrix in a much stronger sense that in the work of Duistermaat \& H\"ormander, i.e. than in Def. \ref{def:Fp}, as it is the generalized inverse of a \emph{Fredholm operator}\footnote{Let us recall that a bounded operator is called Fredholm if the dimension of its kernel and cokernel are finite.}. Furthermore, it  bears much more resemblance to elliptic inverses than the retarded and advanced propagators do, due to its positivity properties \cite{positive}. One can also argue that its definition is a highly canonical one (possibly modulo finite dimensional choices) as it relates directly to the bi-characteristic flow.


Moreover, a recent work of B\"ar and Strohmaier that treats the Dirac operator on compact globally hyperbolic spacetimes with space-like boundary \cite{BS} achieves to set up a Fredholm problem that is in many ways similar to that of Gell-Redman, Haber and Vasy. In B\"ar and Strohmaier's setting, the problem is formulated by imposing boundary conditions that are analogous to Atiyah-Patodi-Singer ones in the Riemannian case. Interestingly, they prove a Lorentzian analogue of the Atiyah-Patodi-Singer theorem \cite{APS1,APS2} and relate the index to quantities of direct physical interest (in particular the so-called chiral anomaly), explaining also the relation to particle creation on curved space-times \cite{BS2} (see also the works of Gibbons \cite{Gi1,Gi2} and Gibbons and Richer \cite{Gi3} for earlier related developments).

\medskip

In the present paper, our main aim is to set up a Fredholm problem on a class of spacetimes similar to that considered in \cite{GHV}, but for the massive Klein-Gordon equation instead of the wave equation (i.e. for $V\neq 0$). On the other hand, we use an approach that is more closely related to the method of \cite{BS} and that in fact can be thought of as its non-compact, infinite time generalization, at least if one disregards distinct features of the Dirac and Klein-Gordon equations. In order to understand better the relation of the so-obtained Feynman propagator with the time-ordered expressions used usually in interacting QFT, we also construct a rather explicit parametrix with  remainder terms that are at both compact and smooth.

\subsection{Main result} We are primarily interested in the class of \emph{as\-ymptotically Min\-ko\-wski spacetimes}, in the sense that $(M,\altg)$ is a Lorentzian manifold (without boundary) with $M=\rr^{1+d}$ and such that:
\[
(\aM)\ \begin{array}{rl}
&\altg_{\mu\nu}(x)- \alteta_{\mu\nu} \in S^{-\delta}_{\std}(\rr^{1+d}), \ \delta>1,\\[2mm]
&(\rr^{1+d}, \altg)\hbox{ is globally hyperbolic},\\[2mm]
&(\rr^{1+d}, \altg) \hbox{ has a  time function }\tilde{t}\hbox{ such that }\tilde{t}-t\in S^{1-\epsilon}_{\std}(\rr^{1+d}),\ \epsilon>0,\\[2mm]
\end{array}
\]
where $\alteta_{\mu\nu}$ is the Minkowski metric and $S_{\std}^{\delta}(\rr^{1+d})$ stands for the class of smooth functions $f$ such that, denoting $\bra x \ket = (1+|x|)^{\12}$,
\[
\p^{\alpha}_{x}f\in O(\langle x\rangle^{\delta- |\alpha|}), \ \alpha\in \nn^{1+d}.
\] 
This way, $\altg$ decays to the flat Minkowski metric simultaneously in time and in the spatial directions in a short-range\footnote{This corresponds to the assumption $\delta>1$.} way. In a similar vein the potential is required to satisfy $\altV(y)- \altm^{2}\in S^{-\delta}_{\std}(\rr^{1+d})$, $\altm>0$. Note that the definition $(\aM)$ covers a similar class of spacetimes to those considered in \cite{BVW,GHV} (the latter are also called asymptotically Minkowski spacetimes therein), but strictly speaking they are not exactly the same: in our setup for instance $(M,\altg)$ is globally hyperbolic, which is not clear from the outset in \cite{BVW,GHV}. 

 \medskip

The main idea in the formulation of the Fredholm problem is to consider `boundary conditions' that select asymptotic data which account for propagation of singularities within only one of the two connected components $\cN^\pm$ of the characteristic set $\cN=\cN^+\cup \cN^-$ of $P$. While in \cite{BS} there is indeed a boundary at finite times (consisting of the union of two time slices), here we need to consider infinite times instead, so boundary conditions are not to be understood literally as they are rather specified at the level of scattering data. 

Let us illustrate how one can separate solutions according to $\cN^\pm$, starting with the example of the free Klein-Gordon operator $P_\free=\p_{t}^{2}- \Delta_{\rx}+ \altm^{2}$. First, for $t,s\in\rr$ let us denote by $\cU_\free(t,s)$ the Cauchy evolution propagator of $P_\free$, i.e. the operator that maps $\cU_\free(t,s):$ \emph{Cauchy data at time $s$} $\mapsto$ \emph{Cauchy data at time $t$}. Then $\cU_\free(t,s)$ is generated by a Hamiltonian which is selfadjoint in the energy space of Cauchy data, and its spectral projections to the half-lines $\rr^\pm$ are given by
\beq\label{eq:cfree}
c_{\free}^{\pm,\vac} \defeq \12\begin{pmatrix}\one & \pm \sqrt{-\Delta_\rx+\altm^2} \\ \pm \sqrt{-\Delta_\rx+\altm^2} & \one\end{pmatrix}.
\eeq
The two operators\footnote{The operators $c_{\free}^{\pm,\vac}$ also have  the interpretation of being the covariances (acting on Cauchy data) of the Minkowski vacuum state, see e.g. the discussion in \cite{inout}.} $c_{\free}^{\pm,\vac}$ project to Cauchy data of solutions propagating with wave front set in $\cN^\pm$; equivalently, this corresponds to splitting the Cauchy evolution in terms of the two groups $\e^{\pm\i t \sqrt{\Delta+m^2}}$. Now with our assumptions, $P$ is close to $P_\free$ at infinity, and therefore it makes sense to use the splitting \eqref{eq:cfree} on the level of scattering data. 

In order to define scattering data in the setting of asymptotically Minkowski spaces we first make a change of variables by means of a diffeomorphism $\chi$ (see Subsect. \ref{s11.1}), which allows to put the metric in the form
\[
\chi^* \altg = -  \altch^{2}(t, \rx)dt^{2}+ \hat  \alth(t, \rx)d\rx^{2},
\]
where $\altch$ tends to $1$ for large $|x|$, while $\hat\alth$ tends to some asymptotic metrics $\hat\alth_{\inout}$ depending on the sign of $t$. In these coordinates, a convenient choice of Cauchy data is $\varrho_s u\defeq (u,-\i\altch^{-1}\p_{t}u)\traa{t=s}$. On the other hand, the natural reference dynamics in this problem (at both future and past infinity) is that of the free Klein-Gordon operator $P_\free$. Let us fix $t=0$ as our reference time. We define the \emph{Feynman} and \emph{anti-Feynman scattering data maps}: 
\[
\bea
\varrho_{\F}&\defeq \slim_{t_\pm\to\pm\infty} \left(c_{\free}^{+,\vac} \cU_{\free}(0,t_+)\varrho_{t_+} + c_{\free}^{-,\vac} \cU_{\free}(0,{t_-})\varrho_{t_-}\right), \\
\varrho_{\rm \overline{F}}&\defeq \slim_{t_\pm\to\pm\infty}\left( c_{\free}^{+,\vac} \cU_{\free}(0,{t_-})\varrho_{{t_-}} + c_{\free}^{-,\vac} \cU_{\free}(0,t_+)\varrho_{t_+}\right),
\eea
\]
as appropriate strong operator limits. We abbreviate  the Sobolev spaces $H^\sobo(\rr^d)$ by $H^\sobo$ and denote by $\cU(t,s)$ the Cauchy evolution propagator of $P$.  Our main result can be stated as follows. 

\begin{theorem}\label{thm:main2}Assume $(\aM)$ and let $m\in\rr$. Consider the Hilbert space 
\beq
\cX_\F^m \defeq  \big\{ u \in (\chi^{-1})^{*} \big(C^{1}(\rr; H^{\sobo+1})\cap C^{0}(\rr; H^{\sobo})\big): \ P u \in \cY^m, \ \varrho_{\aF} u=0\big\},
\eeq
where $\cY^m\defeq(\chi^{-1})^{*}\left(\bra t \ket^{-\gamma} L^2(\rr;H^m)\right)$ and $\textstyle\12<\gamma<\textstyle\12+\delta$. Then $P:\cX_\F^m\to \cY^m$ is Fredholm of index 
\beq\label{eq:theindex2}
\ind P|_{\cX_\F^m\to \cY^m}= \ind (c^{-,\vac}_{\free}W_{\out}^{-1}+ c^{+, \vac}_{\free}W_{\inn}^{-1}),
\eeq
where $W_\outin^{-1}=\lim_{t_\pm\to \pm\infty}\cU_{\free}(0, t_\pm)\cU(t_\pm, 0)$. In particular  the index  is independent on $m$. Furthermore, there exists a Feynman parametrix $G_\F:\cY^m\to\cX_\F^m$ such that $\one-PG_\F$ and $\one-G_\F P$ are  compact operators.
\end{theorem}

Note that  the space  $\cX_{\F}^{m}$ is  a closed subspace of the Hilbert space 
\[
\cX^{m}= \{ u \in (\chi^{-1})^{*} \big(C^{1}(\rr; H^{\sobo+1})\cap C^{0}(\rr; H^{\sobo})\big): \ P u \in \cY^m\}
\]
equipped with the norm    $\| u\|^{2}_{\cX^{m}}= \| \varrho_{0}(\chi^{-1})^{*}u\|_{\cE^{m}}+ \|Pu\|^{2}_{\cY^{m}}$,  where $\cE^{m}$ is the energy space, see 
Def. \ref{defowit}.

As pointed out in \cite{BS}, the condition $\varrho_{\aF}u=0$ can be seen as an analogue of the Atiyah-Patodi-Singer boundary condition (even though this is less evident here as we do not consider the Dirac equation). Furthermore, one could equally well consider the \emph{anti-APS} boundary condition $\varrho_\F u=0$, which leads to an `anti-Feynman' counterpart of Theorem \ref{thm:main2} --- interestingly, just as in \cite{BS}, this differs from the Riemannian case where one boundary condition is preferred over the other. On the other hand, we also prove that the kernel (null space) of $P:\cX_\F^m\to \cY^m$ consists of smooth functions, and that $G_\F$ satisfies a positivity condition $\i^{-1}(G_\F-G_\F^*)\geq0$ reminiscent of the positivity of the spectral measure in the limiting absorption principle. As pioneered in \cite{BS} and \cite{BVW,positive}, this shows once again a striking similarity to the elliptic case.

On the side note it is worth mentioning that our results can also be adapted to asymptotically static space-times with compact Cauchy surface, for in that setting, decay in the spatial directions becomes irrelevant and consequently one can use directly the time decay estimates from \cite{inout} instead of the estimates considered here. 

Furthermore, a large part of the arguments of B\"ar and Strohmaier \cite{BS2} can be applied to our case, and thus allow one to interpret the index of $P:\cX_\F^m\to \cY^m$ in terms of particle creation. 

\subsection{Outline of proofs} An important step in our proof is a procedure derived in detail in \cite{bounded,inout}, which allows us to reduce the whole analysis to the case when the Klein-Gordon operator is of the form
\beq\label{Pintro}
P=\p_t^2 + r(t)\p_t + a(t,\bx,\p_\bx),
\eeq
where $r(t)$ is a multiplication operator and $a(t,\bx,D_\bx)$ is a differential operator with principal symbol $\spexi \cdot \alth_t^{-1}(\bx)\spexi$.

 To account for the decay properties that follow from assumption $(\aM)$, we introduce a pseudodifferential calculus $\Psi^{m,\delta}_\std(\rr;\rr^d)$ ($m,\delta\in\rr$) consisting of pseudodifferential operators with time-depending symbols $a(t, \rx, \spexi)$ satisfying
\[
 \p_{t}^{\gamma}\p_{\rx}^{\alpha}\p_{\spexi}^{\beta}a(t, \rx, \spexi)\in O\big((\langle t\rangle+ \langle\rx\rangle)^{\delta- \gamma- |\alpha|}\langle \spexi\rangle^{m-|\beta|}\big),  \ \gamma\in \nn, \ \alpha, \beta\in \nn^{d}.
\]
Thus, $\delta$ accounts for the asymptotic behaviour for large $|\rx|$ and/or $t$. We also introduce its time-independent counterpart, denoted by $\Psi_{\scc}^{m,\delta}(\rr^{d})$.  We then find that our hypothesis $(\aM)$ implies that one is reduced to the situation covered by the following list of assumptions (keeping in mind that $\delta>1$ is as in $(\aM)$):
\[
(\Hstd) \  \ \beal
r(t)\in \Psi_{\std}^{0, -1-\delta}(\rr; \rr^{d}),\\[2mm]
a(t, \rx, D_{\rx})- a_{\outin}(\rx, D_{\rx})\in \Psi_{\std}^{2, - \delta}(\rr; \rr^{d})  \hbox{ on }\rr^{\pm}\times \rr^d\hbox{ where: }\\[2mm]
a_{\outin}(\rx, D_{\rx})\in \Psi_{\scc}^{2,0}(\rr^{d})\hbox{\ is  elliptic (in the usual } \Psi^2(\rr^d) \hbox{ calculus)} ,\\[2mm]
a_{\outin}(\rx, D_{\rx})= a_{\outin}(\rx, D_{\rx})^{*}\geq C_{\infty}>0.
\eeal  
\]


In this setup, using standard arguments from Fredholm analysis and the well-posedness of the inhomogeneous Cauchy problem
\[
\begin{cases}
Pu=f, \ \ f\in\cY^m \\
\varrho_t u = v, \ \ v\in H^{m+1}(\rr^d)\oplus H^{m}(\rr^d),
\end{cases}
\]
we deduce that the Fredholm property of $P$ acting on the spaces $\cX_\F^m$, $\cY^m$ (strictly speaking their analogue in the reduced setting $(\std)$, so it is not necessary to use a diffeomorphism $\chi$ in the definition) is equivalent to the Fredholm property of the operator
\[
W_\F \defeq W_\out c^{+,\vac}_{\free} +  W_\inn c^{-,\vac}_{\free}\in B(H^{m+1}(\rr^d)\oplus H^m(\rr^d)).
\]
We actually prove the stronger statement that the operators
\beq\label{eq:compa}
W_\F^\dagger W_\F - \one, \ \ W_\F W_\F^\dagger  - \one
\eeq
are \emph{smoothing} and have \emph{decay} properties that imply their compactness (here $^\dagger$ denotes the adjoint with respect to the canonical non-positive charge inner product preserved by the evolution, so that $W_\out^\dagger W_\out=W_\inn^\dagger W_\inn=\one$). The proof is based on the method of approximate diagonalization of the Cauchy evolution, developed successively in the works \cite{junker,JS,GW,GW2,bounded} and improved in the present paper to yield estimates on the decay (in both time and spatial directions) of various terms that account for the relation between the full dynamics and its asymptotic counterparts. The role of this approximate diagonalization, beside providing the basis for the construction of a parametrix, is to split the Cauchy evolution in two parts corresponding to propagation within $\cN^+$ and $\cN^-$, and then our estimates allow us to relate this splitting to the canonical one for the asymptotic dynamics.    

The construction of the Feynman parametrix $G_{\F}$ is based on a formula that is a time-ordered expression\footnote{We mean specifically that its integral kernel is of the form $\theta(t-t')\Lambda^+(x,x')+ \theta(t'-t)\Lambda^-(x,x')$ with $\Lambda^\pm\geq0$.}, but involving only the `diagonalizable' part of the evolution. The evolution being in fact diagonalizable in the sense considered here modulo terms that are smoothing and decaying, this produces a parametrix indeed, and the proof of the wave front set condition follows by standard arguments, detailed previously in \cite{bounded}.

\medskip




It is worth mentioning that compactness of the remainder term in \eqref{eq:compa} was already studied in an analogous problem for the Dirac operator on Minkowski space with external potentials \cite{matsui1,matsui2,BH}, where index formulas have also been derived and the interpretation of the index in terms of particle creation was discussed (see also \cite{BS,BS2}). An interesting topic of further research would thus be to find a short-hand index formula in our setting.

\subsection{Plan of the paper} The paper is organized as follows.

In Sect. \ref{sec2} we introduce the time-dependent pseudodifferential operator classes $\Psi_\std^{m,\delta}$ and state some of their properties. We recall the method of approximate diagonalization of the Cauchy evolution from \cite{bounded,inout}, and we then give a refinement in the setup of assumption $(\Hstd)$ by showing decay of various remainder terms.

In Sect. \ref{sec:abstract} we set up a Fredholm problem for the Klein-Gordon operator, assuming hypothesis $(\Hstd)$, and then we construct a  Feynman parametrix and prove that the remainder terms are compact operators. An important role is played by the approximate diagonalization and the estimates from Sect. \ref{sec2}.

Finally, in Sect. \ref{sec:ams} we consider asymptotically Minkowski spacetimes $(\aM)$. We show that in this case, using the procedure from \cite{bounded,inout} one is reduced to assumption $(\Hstd)$. This allows us to adapt the results from Sect. \ref{sec:abstract} and to prove Thm. \ref{thm:main2}.  As an aside, we show that the retarded and advanced propagators can be obtained as inverses of a bounded operator acting on Hilbert spaces; this gives another analogy to the setting of \cite{GHV}.

Several auxiliary proofs are collected in Appendix \ref{secapp1}. 

\section{Model Klein-Gordon operator}\label{sec2}

\subsection{Notation}\label{sec1.1}The space of differential operators on $\rr^k$ of order $m$ is denoted by $\Diff^m(\rr^k)$. The space of smooth functions  with compact support is denoted $\coinf(\rr^k)$.

From now on, the operator of multiplication by a function $f$ will be denoted by $f$, while the operators of partial differentiation will be denoted by $\pe_{i}$, so that $[\pe_{i}, f]= \p_{i}f$.

%

\subsection{Klein-Gordon operator}\label{ssec:classical} 
In what follows we use the notation $x= (t, \rx)$ for points in $\rr^{1+d}$, $d\geq 1$.

Before considering the Klein-Gordon equation on actual asymptotically Minkowski spacetimes, we first work with a simpler set of assumptions that allow us to write the Klein-Gordon operator in the form
\beq\label{eq:modelP}
 P = \pe_{t}^{2}+ r(t,\rx )\pe_{t}+ a(t, \rx, \pe_{\rx})\in\Diff^2(\rr^{1+d}),
\eeq
where $r(t,\rx )\in \cf(\rr^{1+d})$, and $a(t, \rx, \pe_{\rx})\in\Diff^2(\rr^d)$ depends smoothly on $t$. Specifically, if $\alth(t)$ is a smooth family of smooth Riemannian metrics on $\rr^d$ and $\altV\in\cf(\rr^{1+d})$ is real-valued, we consider the metric
\[
\altg = - dt^{2}+ \alth_{ij}(t,\rx )d\rx^{i}d\rx^{j}
\]
on $\rr^{1+d}$. Then, the Klein-Gordon operator $\tilde P\defeq-\Box_{\altg}+\altV$ equals
\beq\label{eq:modelP2}
 P = |\alth|^{-\12}\pe_{t}|\alth|^{\12}\pe_{t}- |\alth|^{-\12}\pe_{i}\alth^{ij}|\alth|^{\12}\pe_{j}+\altV,
\eeq
which is indeed of the form \eqref{eq:modelP} with
\[
a(t, \rx, \pe_{\rx})= - |\alth|^{-\12}\pe_{i}\alth^{ij}|\alth|^{\12}\pe_{j}+ \altV(t,\rx ) 
\]
and
\[
r(t,\rx )= |\alth|^{-\12}\p_{t}(|\alth|^{\12})(t,\rx ).
\]
In the sense of formal adjoints, the Klein-Gordon operator satisfies $P^*=P$ with respect to the $L^{2}(\rr^{1,d}, |\alth|^{\12}dt d\rx)$
scalar product, and similarly $a(t, \rx, \pe_{\rx})^*=a(t, \rx, \pe_{\rx})$ with respect to the $t$-dependent $L^{2}(\rr^{d}, |\alth|^{\12} d\rx)$ scalar product.

We will often abbreviate $a(t, \rx, \pe_{\rx})$ by $a(t)$ or simply by $a$.

In what follows we introduce the terminology needed to formulate additional assumptions on $a$ and $r$. These will correspond to a scattering situation, i.e. to the case of a metric $\altg$ (resp. potential $\altV$) converging to some asymptotic static metrics $\altg_{\rm out/in}= -dt^{2}+ \alth_{\rm out/in, ij}(\rx)d\rx^{i}d\rx^{j}$ (resp. to some time-independent potentials $\altV_{\outin}$) as $t\to \pm\infty$.


\subsection{Scattering pseudodifferential calculus}\label{scatcat}
We start by introducing a time-dependent pseudodifferential calculus on $\rr^{d}$ that allows to control both Sobolev regularity in the usual sense and decay at space-time infinity, and which is therefore the natural calculus on asymptotically Minkowski spacetimes. 

For $m, \delta\in \rr$  we denote by $S^{m, \delta}_{\std}(\rr; T^{*}\rr^{d})$ the space of smooth functions $a(t, \rx, \spexi)$ such that
\[
 \p_{t}^{\gamma}\p_{\rx}^{\alpha}\p_{\spexi}^{\beta}a(t, \rx, \spexi)\in O((\langle t\rangle+ \langle\rx\rangle)^{\delta- \gamma- |\alpha|}\langle \spexi\rangle^{m-|\beta|}),  \ \gamma\in \nn, \ \alpha, \beta\in \nn^{d}.
\]
The subscript $\std$ refers to the space-time decay properties of the symbols in $(t, \rx)$.
The subspace of symbols which are poly-homogeneous in $\spexi$ will be denoted  by $S^{m, \delta}_{\std, {\rm ph}}(\rr; T^{*}\rr^{d})$. We denote by $\cW_{\std}^{-\infty}(\rr;\rr^{d})$ the space of operator-valued functions $a(t)$ such that 
\[
\|(D_{\rx}^{2}+\rx^{2}+1)^{m}\p_{t}^{n}a(t)(D_{\rx}^{2}+\rx^{2}+1)^{m}\|_{B(L^{2}(\rr^{d}))}\in O(\bra t\ket^{-p}), \ \forall\, m,n,p\in \nn.
\]
These will play the role of error terms in the calculus. Correspondingly, we set:
\[
\Psi^{m, \delta}_{\std}(\rr; \rr^{d})\defeq\Op^{\rm w}(S^{m, \delta}_{\std, {\rm ph}}(\rr; T^{*}\rr^{d}))+ \cW_{\std}^{-\infty}(\rr;\rr^{d}),
\]
where $\Op^{\rm w}$ is the well-known Weyl quantization, defined by
\[
 \Op^{\rm w}(a)u(\rx)= (2\pi)^{-d}\int_{\rr^{2d}} \e^{\i (\rx-\ry)\cdot \spexi}a\big(t,\textstyle\frac{\rx+\ry}{2}, \spexi\big)u(\ry)d\ry  d\spexi.
\]
It is easy to see that
\[
\begin{array}{l}
S^{m, \delta}_{\std, {\rm ph}}(\rr; T^{*}\rr^{d})= (\langle \rx\rangle+ \langle t\rangle)^{\delta}S^{m, 0}_{\std, {\rm ph}}(\rr; T^{*}\rr^{d}),\\[2mm]
\Psi^{m, \delta}_{\std}(\rr; \rr^{d})= (\langle \rx\rangle+ \langle t\rangle)^{\delta}\Psi^{m, 0}_{\std}(\rr;\rr^{d}),
\end{array}
\]
which allows us to reduce ourselves to the case $\delta=0$.

Omitting the variable $t$ in the above definitions,  we also obtain classes of (time-independent) symbols and pseudodifferential operators on $\rr^{d}$, which will be denoted respectively by $S^{m, \delta}_{\rm sd}(T^{*}\rr^{d})$, $\Psi^{m, \delta}_{\rm sd}(\rr^{d})$ and $\cW^{-\infty}_{\sd}(\rr^{d})$, where the subscript $\sd$ refers to space decay properties of the symbols or operators.  

The classes $\Psi^{m, \delta}_{\sd}(\rr^{d})$ are the well-known `scattering  pseudodifferential operators', see e.g. \cite{cordes,parenti,moreshubin}. The error terms $\cW^{-\infty}_{\sd}(\rr^{d})$ can be described as
\beq\label{eq:descr}
\cW^{-\infty}_{\sd}(\rr^{d})= \textstyle\bigcap_{m\in\rr}B\big(\langle \rx\rangle^{m}H^{-m}(\rr^d),\langle \rx\rangle^{-m}H^m(\rr^d)\big),
\eeq
where $H^m(\rr^d)$ is the standard Sobolev space of order $m$. Using \eqref{eq:descr} we equip $\cW_{\sd}^{-\infty}(\rr^{d})$ with its canonical Fr\'echet space topology; this is needed to define in the obvious way spaces such as $\cf(\rr^2;\cW_{\sd}^{-\infty}(\rr^{d}))$. We will also occasionally need the broader class of smoothing operators $\cW^{-\infty}(\rr^d)$, defined as in \eqref{eq:descr} but with the $\langle \rx\rangle^{\pm m}$ weights ommitted. 
 
We will need an appropriate notion of \emph{ellipticity} and principal symbol for the $\Psi^{m, 0}_{\rm sd}(\rr^{d})$ and  $\Psi_{\std}^{m, 0}(\rr; \rr^{d})$ classes. For our purposes, elliptic operators in $\Psi^{m, 0}_{\rm sd}(\rr^{d})$ will be simply those which are elliptic in the usual sense\footnote{Thus, we do \emph{not} consider here ellipticity in the sense of the scattering pseudo\-differential calculus \cite{cordes,parenti,moreshubin}.} in $\Psi^{m}(\rr^{d})$, i.e. $a\in\Psi^{m}(\rr^{d})$ is elliptic if there exists $C>0$ such that
\beq\label{eq:elliptic}
| \sigma_{\rm pr}(a)|\geq C |\spexi|^{m}, \ \ |\spexi|\geq 1,
\eeq
where $\sigma_{\rm pr}(a)$ is the principal symbol of $a$ defined in the usual way, see e.g. \cite{shubin}. Furthermore, $a\in \Psi^{m, \delta}_{\std}(\rr; \rr^{d})$ is said to be elliptic if $a(t)$ is elliptic for all $t\in\rr$ and the constant $C$ in (\ref{eq:elliptic}) is uniform in $t$.

Let us remark that the $\Psi_{\std}^{m,\delta}(\rr;\rr^d)$ pseudodifferential calculus has structural properties very analogous to the $\Psi_{\td}^{m,\delta}(\rr;\rr^d)$ calculus introduced in \cite{inout}, which is defined by omitting $\bra x \ket$ in the relevant definitions (so that only decay in time is taken into account). As a consequence, many results from \cite{inout} can be adapted to the present setup. In particular, a variant of Seeley's theorem is  valid for the $\Psi_{\std}^{m, 0}(\rr; \rr^{d})$ classes; it is proved in complete analogy to \cite[Thm. 3.7]{inout} by  reduction to the general framework of \cite{alnv1}, see also the arguments in \cite[Subsect. 5.3]{bounded}.

\begin{theorem}\label{seeley-std}
 Let $a\in\Psi_{\std}^{m, 0}(\rr; \rr^{d})$ be elliptic, selfadjoint with $a(t)\geq c_{0}\one$ for some $c_{0}>0$. Then for any $\alpha\in \rr$, $a^{\alpha}\in\Psi_{\std}^{\alpha m, 0}(\rr; \rr^{d})$  and $\sigma_{\rm pr}(a^{\alpha})(t)= \sigma_{\rm pr}(a(t))^{\alpha}$.
\end{theorem}
\proof We will simply outline the main steps of the proof, which consists in verifying the abstract conditions in \cite{alnv1}. We use the notations in  \cite[Subsect. 5.3]{bounded}. 

We choose as Hilbert space $\cH= L^{2}(\rr_{t}\times \rr^{d}_{\rx})$.   As injective operator  on $\cH$ belonging to $\cW_{\rm std}(\rr; \rr^{d})$ we choose $R= \e^{- (D_{\rx}^{2}+ \rx^{2}+ t^{2}+1)}$.
It is easy to see that if $a\in S^{m, 0}_{\std, {\rm ph}}(\rr; T^{*}\rr^{d})$ and $\Op(a)\in \cW_{\std}^{-\infty}(\rr; \rr^{d})$ then $a\in S^{-\infty, 0}_{\std, {\rm ph}}(\rr; T^{*}\rr^{d})$.
It remains to check the spectral invariance property of $\cW_{\std}^{-\infty}(\rr; \rr^{d})$, ie to prove that if $R\in \cW_{\std}^{-\infty}(\rr; \rr^{d})$ and $\one -R$ is boundedly invertible on $\cH$, then $(\one - R)^{-1}\in \one + \cW_{\std}^{-\infty}(\rr; \rr^{d})$. We write $\cH$ as $L^{2}(\rr_{t}; L^{2}(\rr^{d}))$ and $R= \int^{\oplus}_{\rr}R(t)dt$, hence  $(\one - R)^{-1}= \int_{\rr}^{\oplus}(\one - R(t))^{-1}dt$. As in \cite[Lem. 5.5]{bounded}, $\one - R(t)$ is boundedly invertible on $L^{2}(\rr^{d})$ for all $t\in \rr$ and we have
\[
(\one - R(t))^{-1}= \one + R(t)+ R(t)(\one - R(t))^{-1}R(t)\defeq\one+ R_{1}(t).
\]
 Using that $\p_{t}(\one - R(t))^{-1}= (\one - R(t))^{-1}\p_{t}R(t)(\one - R(t))^{-1}$ and Leibniz rule we obtain that $R_{1}(t)\in \cW_{\rm std}(\rr; \rr^{d})$. \qed

Later on we will need the following auxiliary result on fractional powers of elliptic operators, the proof of which  is deferred to Subsect. \ref{ssecap1}.
\begin{proposition}\label{l5.1}
 Let   $a_{i}\in\Psi_{\std}^{2,0}(\rr; \rr^d)$ ($i=1,2$) be elliptic, with $a_{i}=a_{i}^{*}$ and $a_{i}(t)\geq c_{0}\one$ for some $c_{0}>0$. Suppose that $a_{1}- a_{2}\in \Psi_{\std}^{2, -\delta}(\rr; \rr^d)$ with $\delta>0$. Then  for each $\alpha\in \rr$ one has:
 \[
 a_{1}^{\alpha}- a_{2}^{\alpha}\in \Psi_{\std}^{2\alpha, -\delta}(\rr;\rr^d).  
 \]
 \end{proposition}

Let us now consider the ressummation of symbols.  We denote 
\[
\Psi^{-\infty,-\delta}_{\std}(\rr; \rr^d)\defeq\textstyle\bigcap_{m\in\rr}\Psi^{m,-\delta}_{\std}(\rr; \rr^d),
\]
which is a subclass of (but is not equal to) $\cW_{\std}^{-\infty}(\rr;\rr^{d})$. The lemma below is obtained by exactly the same arguments as in \cite[Lem. 3.11]{inout}.

\begin{lemma}\label{l5.2}
Let $\delta\in \rr$ and let $(m_{j})$ be a real sequence decreasing to $-\infty$. Then
if  $a_{j}\in \Psi_{\std}^{ m_{j},-\delta}(\rr; \rr^d)$ there exists $a\in \Psi_{\std}^{ m_{0},-\delta}(\rr; \rr^d)$, unique modulo $\Psi^{-\infty,-\delta}_{\std}(\rr; \rr^d)$, such that
 \[
 a\sim \sum_{j=0}^{\infty}a_{j}, \hbox{ i.e. } \forall N\in \nn, \ 
 a-\sum_{j=0}^{N}a_{j}\in\Psi_{\std}^{ m_{N+1},-\delta}(\rr; \rr^d).
  \]
 \end{lemma}

Thus when performing the ressumation of $(a_{j})$ we have control of the decay of the error terms (as they belong to $\Psi^{-\infty,-\delta}_{\std}(\rr; \rr^d)$).

\subsection{Assumptions of the model}\init\label{scassumption} We are now ready to state the precise assumptions on the model Klein-Gordon operator 
\[
 P = \pe_{t}^{2}+ r(t,\rx )\pe_{t}+ a(t, \rx, \pe_{\rx}).
\]
Namely, we assume that there exist $\delta>0$, $a_{\outin}(\rx, D_{\rx})\in \Psi_{\scc}^{2,0}(\rr^{d})$ and a constant $C_\infty$ such that:
\[
(\Hstd) \  \ \beal
a(t, \rx, D_{\rx})= a_{\outin}(\rx, D_{\rx})+ \Psi_{\std}^{2, - \delta}(\rr; \rr^{d})  \hbox{ on }\rr^{\pm}\times \rr^d,\\[2mm]
r(t)\in \Psi_{\std}^{0, -1-\delta}(\rr; \rr^{d}),\\[2mm]
a_{\outin}(\rx, D_{\rx})\in \Psi_{\scc}^{2,0}(\rr^{d})\hbox{\ is  elliptic},\\[2mm]
a_{\outin}(\rx, D_{\rx})= a_{\outin}(\rx, D_{\rx})^{*}\geq C_{\infty}>0.
\eeal  
\]
Note that this gives a setup which is a particular case of that considered recently  in \cite{inout}. Here we assume decay of various terms both in time and in the spatial variables, while in \cite{inout} only time decay is required (and more general manifolds are allowed in the place of $\rr^d$).

\subsection{Approximate diagonalization of the Cauchy evolution}\label{sec2.3} For $t\in\rr$ let us denote by $\varo_{t}$ the Cauchy data map
\[
\bea
\varo_t :  \cinf(\rr; \cD'(\rr^{d}))&\to \cD'(\rr^{d})\oplus \cD'(\rr^{d})\\
u&\mapsto  (u(t), \i^{-1}\p_{t}u(t)).
\eea
\]
Furthermore, for $t,s\in\rr$ let us denote by $\cU(t,s)$ the \emph{Cauchy evolution propagator} for $P$, i.e., $\cU(t,s)v$ is by definition the  Cauchy data $\varrho_t u$ of the unique solution of the Cauchy problem
\[
\begin{cases}
Pu =0,\\
\varrho_s u =v.
\end{cases}
\]
The main ingredient in our approach is a refined version of the approximate diagonalization of $\cU(t,s)$ developed in \cite{bounded,inout} in the context of Hadamard states. In what follows we recall its outcome and discuss how it can be improved in the setup of assumption $(\std)$. 

The first step consists in performing an approximate factorization of the Klein-Gordon operator $P=\pe_{t}^{2}+ r \pe_{t}+ a$. This means specifically that one constructs a family of elliptic pseudo\-dif\-fe\-rential operators $b^+(t)$, smoothly depending on $t$, and such that
\begin{equation}
\label{eq-e1}
P=(\pe_{t}+ \i b^{+}(t)+ r(t))\circ (\pe_{t}- \i b^{+}(t))+r_{\infty}^{+}(t),
\end{equation}  
where $r_{\infty}^{+}(t)$ is a smoothing error. Once this is done, one actually gets immediately a second solution to this problem by
setting 
\beq\label{eq:adalk}
b^{-}= - (b^+)^{*}, \ \ r_{\infty}^{-}=(r_{\infty}^+)^{*}.
\eeq
Indeed, by taking the adjoint of both sides of (\ref{eq-e1}) with respect to the $t$-dependent inner product $L^{2}(\rr^d, |\alth|^{\12}d\rx)$, and using that
 \[
\bea
( \p_{t}b^+)^{*}&= \p_{t}((b^+)^{*})+ r (b^+)^{*}- (b^+)^{*}r\\ &=-\p_{t}b^{-} - r b^{-}+ b^- r,
\eea
 \]
one obtains that
\[
P=(\pe_{t}+ \i b^{-}(t)+ r(t))\circ (\pe_{t}- \i b^{-}(t))+r_{\infty}^{-}(t).
\] 
The pair of identities
\begin{equation}
\label{eq-e1er}
P=(\pe_{t}+ \i b^{\pm}(t)+ r(t))\circ (\pe_{t}- \i b^{\pm}(t))+r_{\infty}^{\pm}(t).
\end{equation}  
is then used to re-express the Cauchy evolution of $P$ in terms of the Cauchy evolution of $\pe_{t}- \i b^{+}(t)$ and $\pe_{t}- \i b^{-}(t)$.



In Prop. \ref{p5.1} below, we state our result on the existence of $b^\pm(t)$ as above and satisfying in addition decay estimates that are crucial for the proofs in the rest of the paper. To simplify notation we write $b_1(t)=b_2(t)+ \Psi_{\std}^{m, \delta}(\rr^{\pm}; \rr^d)$ if $b_1(t)$, $b_2(t)$ are two $t$-dependent operators such that  $b_1(t)-b_2(t)\in \Psi_{\std}^{m, \delta}(\rr; \rr^d)$ on $\rr^{\pm}\times \rr^d$. By an argument from \cite{inout} there exists $\varphi\in C_{\rm c}^\infty(\rr)$ such that $a(t)+ \varphi(a(t))>0$ and $a(t)+ \varphi(a(t)) \sim a_{\outin}$ on $\rr^\pm$ in the sense that 
\[
\exists\, c>0 \, \hbox{ s.t. } \, c^{-1} a_{\outin}\leq a(t)+ \varphi(a(t)) \leq c\, a_{\outin}  \hbox{ on } \rr^\pm.
\]
Moreover, $\varphi(a(t))\in\cf_{\rm c}(\rr;\cW_\std^{-\infty}(\rr^d))$.
We then set $\epsilon(t)\defeq \big(a(t)+\varphi(a(t))\big)^\12$ and $\epsilon_\outin\defeq a_\outin^\12$.


\begin{proposition} \label{p5.1} There exist $b^\pm(t)= \pm\epsilon(t)+ \Psi_{\std}^{0,-1-\delta}(\rr; \rr^d)$ that solve \eqref{eq:adalk} and \eqref{eq-e1er} with $r_{-\infty}^{\pm}\in \Psi^{-\infty, -1- \delta}_{\std}(\rr; \rr^d)$, and that satisfy
\beq
b^\pm(t)=\pm\epsilon_{\outin}+ \Psi^{1,-\delta}_{\std}(\rr^{\pm}; \rr^d).
\eeq
Moreover, $b^\pm(t)$ can be chosen in such way that
\beq\label{eq:addi}
(b^+(t)- b^{-}(t))^{-1}\geq C(t) \epsilon(t)^{-1}
\eeq
holds for some $C(t)>0$.
\end{proposition}

The proof is completely analogous to \cite{inout}, using Prop. \ref{l5.1} instead of \cite[Prop. 3.10]{inout}.

The approximate diagonalization of $\{\cU(t,s)\}_{t,s\in\rr}$ proceeds now as follows. We set
\beq\label{eq:defT}
\bea
&T(t)\defeq  \i^{-1}\mat{\one}{-\one}{b^{+}}{-b^{-}}(b^{+}- b^{-})^{-\12}, \\
&T^{-1}(t)= \i (b^{+}- b^{-})^{-\12}\mat{-b^{-}}{\one}{-b^{+}}{\one},
\eea
\eeq
which is well defined by \eqref{eq:addi}. We then define a new evolution group $\{\cU^{\adg}(t,s)\}_{t,s\in\rr}$ by
\beq\label{intolo}
\cU(t,s)\eqdef T(t)\cU^{\adg}(t,s)T(s)^{-1}. 
\eeq
We denote by $H(t)$ the generator of $\{\cU(t,s)\}_{t,s\in\rr}$, defined by 
\beq\label{eq:defHa}
\frac{\p}{\p t}\cU(t,s)\eqdef \i  H(t)\cU(t,s),
\eeq
and similarly we define $H^{\adg}(t)$, the generator of $\{\cU^{\adg}(t,s)\}_{t,s\in\rr}$. Explicitly, the former equals
\[
H(t)=\mat{0}{\one}{a(t)}{\i r(t)}.
\] 
By a direct computation one finds that the latter is of the form
\beq\label{eq:defH}
H^{\adg}(t)=H^{\dg}(t)-V^{\adg}_{-\infty}(t),
\eeq
where $H^{\dg}(t)$ is a diagonal matrix of pseudodifferential operators (smoothly depending on $t$) and $V^{\adg}_{-\infty}(t)$ is a smoothing, decaying remainder. More precisely,
\beq\label{eqdefkl}
H^{\dg}(t)= \mat{\epsilon^{+}}{0}{0}{\epsilon^{-}},
\eeq
where the components $\epsilon^\pm(t)$ are
\beq\label{eq:rbpm}
\bea
\epsilon^\pm &= - b^{\mp}+\i r+ [(b^{+}- b^{-})^{-\12}, b^{\mp}]-\i \p_{t}(b^{+}- b^{-})^{-\12}(b^{+}- b^{-})^{\12}\\
& \phantom{=\,} - (b^{+}- b^{-})^{-\12}r_{-\infty}^{\mp} (b^{+}- b^{-})^{-\12},
\eea
\eeq  
and furthermore,
\beq\label{eq:defV}
\bea
V^{\adg}_{-\infty}(t)&= (b^{+}- b^{-})^{-\12}\mat{r_{-\infty}^{-}}{-r_{-\infty}^{-}}{r_{-\infty}^{+}}{-r_{-\infty}^{+}}(b^{+}- b^{-})^{-\12}\\
&\phantom{=- (b^{+}- b^{-})^{-\12}}\in \Psi^{-\infty, -1- \delta}_{\std}(\rr; \rr^{d})\otimes B(\cc^{2}),
\eea
\eeq
where $r^\pm_{-\infty}\in\Psi^{-\infty, -1- \delta}_{\std}(\rr; \rr^d)$ are the remainder terms from \eqref{eq-e1er}. This way, 
the evolution $\cU^{\dg}(t,s)$ generated by $H^{\dg}(t)$ is diagonal, 
and moreover, as shown in \cite{bounded}:
\begin{equation}
\label{turlututu}
\bea
\cU(t,s)&=T(t)\cU^{\adg}(t,s)T(s)^{-1}\\
&= T(t)\cU^{\dg}(t,s)T(s)^{-1}+ \cinf(\rr^{2}; \cW^{-\infty}(\rr^d)).
\eea
\end{equation}

\begin{proposition}\label{propoesti}
 Assume $(\std)$ and let $\epsilon^{\pm}(t)$ be as defined in \eqref{eq:rbpm}. Then
\begin{equation}
\label{e11.50}
\begin{array}{l}
\epsilon^{\pm}(t)+b^{\mp}(t) \in  \Psi^{0,-1- \delta}_{\std}(\rr; \rr^{d}),\\[2mm]
\epsilon^{\pm}(t)\mp \epsilon(t)\in  \Psi^{0, -1- \delta}_{\std}(\rr; \rr^{d}),
\end{array}
\end{equation}
and consequently,
\begin{equation}
\label{e11.4}
H^{\adg}(t)= \mat{\epsilon(t)}{0}{0}{-\epsilon(t)}+ \Psi^{0, -1- \delta}_{\std}(\rr; \rr^{d})\otimes B(\cc^{2}).
\end{equation} 
 \end{proposition}
\proof From the definition of $\epsilon^\pm$ and the fact that $r_{-\infty}^{\pm}\in\Psi^{-\infty, -1- \delta}_{\std}(\rr; \rr^d)$ one obtains that $\epsilon^\pm = -b^\mp+r_b^\mp$, where
\[
r_b^\pm=\i r+ [(b^{+}- b^{-})^{-\12}, b^{\pm}]-\i \p_{t}(b^{+}- b^{-})^{-\12}(b^{+}- b^{-})^{\12}.
\]
Thus, to get the first part of \eqref{e11.50} it suffices to prove that $r_b^\pm \in\Psi^{0, -1- \delta}_{\std}(\rr; \rr^d)$. This can be performed in exact analogy to \cite{inout} using Seeley's theorem and composition properties of the $\Psi_\std^{m,\delta}(\rr;\rr^d)$ calculus.

The second part of \eqref{e11.50} then follows using Prop. \ref{p5.1}. The last statement is a direct consequence of \eqref{e11.50} and \eqref{eq:defH}--\eqref{eq:rbpm}.\qed


\subsection{Symplectic properties of the approximate diagonalization}

It is well known that there is a symplectic form preserved by the Cauchy evolution of $P$. In our setup this can be written as the identity
\beq\label{eq:conserveq}
\cU(t,s)^*q\cU(t,s)=q, \hbox{ where }  q\defeq \mat{0}{\one}{\one}{0},
\eeq
for all $t,s\in\rr$. The operators $T(t)$ are defined in such way that 
\[
T^{*}(t)q T(t)= q^{\adg}, \hbox{ where } q^{\adg}\defeq\mat{\one}{0}{0}{-\one},
\]
and consequently we have for the approximately diagonalized evolution an analogue of \eqref{eq:conserveq}, namely
\beq\label{eq:conserveq2}
\cU^\adg(t,s)^*q^\adg\cU^\adg(t,s)=q^\adg.
\eeq
On the level of the respective generators, this translates to
\beq\label{eq:Hadjo}
H(t)^{*}q= q H(t), \ \ H^{\adg}(t)^{*}q^{\adg}= q^{\adg} H^{\adg}(t).
\eeq
Using the latter we can deduce that $H^{\dg}(t)= H^{\dg*}(t)$, and hence $\epsilon^{\pm}(t)^*=\epsilon^{\pm}(t)$.

It is convenient to denote by $A^\dag$ the adjoint with respect to the inner product defined by either $q$ or $q^\adg$ whenever it is clear from the context which of the two is meant. For instance, this way \eqref{eq:Hadjo} can be rewritten as $H(t)^\dag=H(t)$ and $H^{\adg}(t)^\dag=H^{\adg}(t)$, though one has to keep in mind that the former refers to $q$ and the latter to $q^\adg$. 

\section{Feynman inverses from scattering data in the model case}\init\label{sec:abstract}

\subsection{Setup} In this section we consider again the model Klein-Gordon operator studied in Subsect. \ref{sec2.3}:
\beq\label{eq:defPstd}
P= \pe_{t}^{2}+ r(t, \rx)\pe_{t}+ a(t, \rx, \pe_{\rx}),
\eeq
and  denote by $P_{\outin}$  the asymptotic Klein-Gordon operators
\[
P_{\outin}\defeq \pe_{t}^{2}+ a_{\outin}(\rx, \pe_{\rx}).
\]
We will assume conditions $(\Hstd)$ with $\delta>1$, which corresponds to a {\em short-range} situation. 

By a parametrix for $P$ we will mean an operator $G_I$ such that $PG_I-\one$ and $G_I P-\one$ have smooth Schwartz kernel. Duistermaat and H\"ormander proved the existence of a Feynman parametrix $G_\F$, or parametrix with \emph{Feynman type wave front set}, i.e.
\[
\WF'(G_\F)= (\diag_{T^*M})\cup\textstyle\bigcup_{s\leq 0}(\Phi_s(\diag_{T^*M})\cap \pi^{-1}\cN).
\]
This means that up to singularities on the full diagonal $\diag_{T^*M}$ of $T^*M\times T^*M$, $\WF'(G_\F)$ is contained in the backward flowout of $\diag_{T^*M}$ by the bicharacteristic flow (here acting on the left component of $T^*M\times T^*M$, accordingly $\pi$ is the projection to that component).
 
Our primary goal will be to prove that for suitably chosen Hilbert spaces of distributions $\cX_I^m, \cY^m$, the operator $P:\cX_I^m\to\cY^m$ is Fredholm, i.e. its kernel and cokernel are of finite dimension. This guarantees the existence of pseudo-inverses, i.e. operators $G_I:\cY^m \to \cX_I^m$ such that $P G_I-\one$ and $G_I P - \one$ are compact. 

We will be interested in constructing a pseudo-inverse that is at the same time a Feynman parametrix. This will be based on the reduction to the almost diagonalized dynamics $\cU^{\adg}(t,s)$ introduced in  Subsect. \ref{sec2.3}.

 \subsection{Notation} First, we introduce some further notation needed to relate in an efficient way various objects related to the different dynamics.
As a rule, all objects related to the almost diagonalized situation are decorated with a superscript $\adg$. We recall that $L^{2}(\rr^{1+d})$ is equipped with the  scalar product
\[
(u|v)\defeq\int \bar{u}v|h_{t}|^{\12}dtd\rx.
\]
\subsubsection{Operators}
Let us recall that  the operators $\AH(t)$, $H^{\adg}(t)$, $T(t)$ are defined respectively in \eqref{eq:defHa}, \eqref{eq:defH}, \eqref{eq:defT}.   

\begin{notations}
\item We set for $u\in \cinf(\rr; \cD'(\rr^{d}))$, $u^{\adg}\in \cinf(\rr; \cD'(\rr^{d})\oplus \cD'(\rr^{d}))$:
\[
\begin{array}{l}
\varo_{t}u= (u(t), \i^{-1}\p_{t}u(t)), \ \ \varo^{\adg}_{t}u^{\adg}\defeq u^{\adg}(t),\\[2mm]
(Tu^{\adg})(t)\defeq T(t)u^{\adg}(t), \ \ (\varo u)(t)\defeq  \varo_{t}u(t).
\end{array}
\]
 Setting also $\pi_{i}(u_{0}, u_{1})= u_{i}$ and using the standard notation  $D_{t}= \i^{-1}\pe_{t}$, we have:
\begin{equation}
\label{e100.-1}
P= - \pi_{1} (D_{t}- \AH(t))\varrho.
\end{equation}
\item We set:
\[
P^{\adg}\defeq  D_{t}- H^{\adg}(t),
\]
and an easy computation shows that:
\beq\label{e100.0}
TP^{\adg}T^{-1}= D_{t}- \AH(t), \hbox{\ hence }P= -\pi_{1}TP^{\adg}T^{-1}\varo. 
\eeq

\item We  denote by $\AH_{\outin},H^{\adg}_{\outin},T_{\outin}$, the analogues of $\AH(t),H^{\adg}(t), T(t)$ with $a(t), r(t)$ replaced by $a_{\outin}, 0$.

\item The Cauchy evolutions generated by  $\AH(t)$, $\AH_{\outin}$, $H^{\adg}(t)$, $H^{\adg}_{\outin}$ are denoted by $\cU(t,s)$,  $\cU_{\outin}(t,s)$, $\cU^{\adg}(t,s)$, $\cU_{\outin}^{\adg}(t,s)$. We recall that:
\beq\label{e100.10}
\cU(t,s)= T(t)\cU^{\adg}(t,s)T^{-1}(s),  \ \ \cU_{\outin}(t,s)= T_{\outin}\cU^{\adg}_{\outin}(t,s)T^{-1}_{\outin}.
\eeq
We also recall that $\cU_{(\outin)}^{(\adg)}(t,s)$, $\cU_{(\outin)}^{(\adg)}(t,s)$ are symplectic for $q^{(\adg)}$.
\end{notations}

\begin{notations}
\item Operators of the form $a\otimes \one_{\cc^{2}}$  will often be abbreviated by $a$ for simplicity.  
\end{notations}

\subsubsection{Function spaces}
 We will abbreviate by $H^{\sobo}$ the Sobolev spaces $H^{\sobo}(\rr^{d})$.

\begin{notations} 
\item Furthermore, we set:
 \[
\cE^{\sobo}\defeq H^{\sobo+1}\oplus H^{\sobo}, \ \ \cH^{\sobo}\defeq H^{\sobo}\oplus H^{\sobo}, \ \sobo\in \rr.
\]
As usual we  define $\cE^{ \infty}\defeq \bigcap_{\sobo\in \rr}\cE^{\sobo}$, $\cE^{-\infty}\defeq \bigcup_{\sobo\in \rr}\cE^{\sobo}$ and similarly for $\cH^{\infty}$, $\cH^{-\infty}$, equipped with their canonical topologies.
\item If $\cE$ is a Banach space, $k\in \nn$, we denote by $C^{k}(\rr; \cE)$ the Banach space of $\cE-$valued functions with norm
\[
\| u\|_{C^{k}(\rr; \cE)}= \sum_{0\leq l\leq k}\sup_{t\in \rr}\| \p_{t}^{l}u(t)\|_{\cE}.
\]
\end{notations}

We will frequently use  the fact that  $T(t):  \cH^{\sobo+\12}\to \cE^{\sobo}$ is boundedly invertible with $\|T(t)\|, \|T^{-1}(t)\|$ uniformly bounded in $t$.  From \cite[Prop. 5.6]{inout} we know that:
\begin{equation}
\label{e11.51}
\sup_{t,s\in\rr}\|\cU^{\adg}(t,s)\|_{B(\cH^{\sobo})}<\infty, \ \ \sup_{t,s\in \rr}\| \cU(t,s)\|_{B(\cE^{\sobo})}<\infty.
\end{equation}

\subsection{M{\o}ller (wave) operators}\label{orlov} We now introduce the relevant objects from scattering theory. We will consider $t=0$ as our fixed reference time. It is a standard fact, derived using \eqref{e11.4}, \eqref{e11.50}  and the so-called Cook argument (see e.g. \cite{dergersc}), that the M{\o}ller operators
\beq\label{azer}
W^{\adg}_\sca \defeq \lim_{t\to\pm\infty} \cU^{\adg}(0,t)\cU^{\adg}_{\sca}(t,0 )\in B(\cH^{\sobo})
\eeq
exist and are invertible with inverses given by
\beq\label{azor}
(W^{\adg}_\sca)^{-1} = (W_\sca^{\adg})^\dagger=\lim_{t\to\pm\infty} \cU^{\adg}_{\outin}(0,t)\cU^{\adg}(t,0)\in B(\cH^{\sobo}).
\eeq
Using then \eqref{e100.10} and the fact that $T^{-1}(t)T_{\outin}- \one$ tends to $0$ in  $B(\cH^{\sobo})$ when $t\to \pm\infty$, we obtain the existence of
\begin{equation}
\label{e100.11}
W_{\sca}\defeq \lim_{t\to \pm\infty}\cU(0,t)\cU_{\sca}(t,0)\in B(\cE^{\sobo}),
\end{equation}
with inverses
\beq\label{e100.12}
(W_\sca)^{-1} = (W_\sca)^\dagger=\lim_{t\to\pm\infty} \cU_{\outin}(0,t)\cU(t,0)\in B(\cE^{\sobo}),
\eeq
and satisfying the identities
\begin{equation}
\label{e100.12b}
W_{\outin}= T(0)W^{\adg}_{\outin}T^{-1}_{\outin}.
\end{equation}
\begin{remark}\label{remstupid}
 Strictly speaking  $W^{(\adg)}_{\outin}$ acting on $\cE^{\sobo}$ or $\cH^{\sobo}$ should be denoted by, e.g., $W_{\outin}^{(\adg), \sobo}$ to indicate its dependence on $\sobo$. However since $W_{\outin}^{(\adg), \sobo}$ is the closure of $W_{\outin}^{(\adg), \sobo'}$ for any $\sobo'>\sobo$, we will  often dispense with the exponent $\sobo$ in the sequel. The same remark applies to $(W_{\outin}^{(\adg)})^{-1}$.
\end{remark}
\subsection{Compactness properties of $W^{\adg}_{\outin}$} 
Let us denote by $\pi^\pm$ the projections
\beq\label{defdepiplusmoins}
\pi^{+}= \mat{\one}{0}{0}{0}, \ \ \pi^{-}=\mat{0}{0}{0}{\one}.
\eeq
The decay properties implied by assumption $(\Hstd)$ have the following important consequence. 

\begin{proposition}\label{prop:stdcase} Assume condition $(\Hstd)$ with $\delta>1$ and let $\alpha<\delta/2$. Then 
\[
W^{\adg}_\outin \pi^+(W^{\adg}_\outin)^{-1}- \pi^+ \in \langle\rx\rangle^{-\alpha}\cW^{-\infty}(\rr^{d})\langle\rx\rangle^{-\alpha}\otimes B(\cc^{2}).
\]
It follows that $[W^{\adg}_{\outin}, \pi^{+}]$ is a compact operator on   $\cH^{\sobo}$ for $\sobo\in \rr$.
\end{proposition}

To prove Prop. \ref{prop:stdcase}, we will need the following lemma, the proof of which is given in Appendix \ref{apota}.
\begin{lemma}\label{turlututi}
 Assume conditions $(\Hstd)$ for $\delta>0$. Then for all $m, k\in \rr^{+}$:
 \[
\sup_{t\geq 0}\|\langle D_{\rx}\rangle^{m}\langle \rx\rangle^{k}\cU^{\adg}(0,t)(\langle \rx\rangle +\langle t\rangle)^{-k}\langle D_{\rx}\rangle^{-m}\|_{B(\cH^{0})}<\infty.
\]
\end{lemma}

{\bf\noindent Proof of Prop. \ref{prop:stdcase}.}  Let us  set  $W^{\adg}_{\outin}(t)= \cU^{\adg}(0,t)\cU^{\adg}_{\outin}(t, 0)$. We have for $m\in \nn$, $\alpha>0$:
\beq\label{taratata}
\begin{array}{l}
\langle D_{\rx}\rangle^{m}\langle\rx\rangle^{\alpha}\p_{t}(W^{\adg}_{\outin}(t)\pi^+ W^{\adg}_{\outin}(t)^{-1})\langle\rx\rangle^{\alpha}\langle D_{\rx}\rangle^{m}\\[2mm]
=\langle D_{\rx}\rangle^{m}\langle\rx\rangle^{\alpha}\cU^{\adg}(0,t)[R_{-\infty}(t), \pi^+]\cU^{\adg}(t,0)\langle\rx\rangle^{\alpha}\langle D_{\rx}\rangle^{m}\\[2mm]
=\langle D_{\rx}\rangle^{m}\langle\rx\rangle^{\alpha}\cU^{\adg}(0,t)(\langle\rx\rangle+ \langle t\rangle)^{-\alpha}\langle D_{\rx}\rangle^{-m}\\[2mm]
\phantom{=}\times \langle D_{\rx}\rangle^{m}(\langle\rx\rangle+ \langle t\rangle)^{\alpha}[R_{-\infty}(t), \pi^+](\langle\rx\rangle+ \langle t\rangle)^{\alpha}\langle D_{\rx}\rangle^{m}\\[2mm]
\phantom{=}\times \langle D_{\rx}\rangle^{-m}(\langle\rx\rangle+ \langle t\rangle)^{-\alpha}\cU^{\adg}(t,0)\langle\rx\rangle^{\alpha}\langle D_{\rx}\rangle^{m}\\[2mm]
\eqdef  R_{m, \alpha}(t)\times M_{m, \alpha}(t)\times R_{m, \alpha}(t)^{\dag}.
\end{array}
\eeq

Since $R_{-\infty}(t)\in \Psi^{-\infty, -1- \delta}_\std(\rr; \rr^d)\otimes B(\cc^{2})$ we know that 
\[
\|M_{m, \alpha}(t)\|_{B(\cH^{0})}\in O(\langle t\rangle^{-1- \delta+ 2\alpha}).
\] 
From Lemma \ref{turlututi} we know that  $\|R_{m, \alpha}(t)\|_{B(\cH^{0})}\in O(1)$, which implies the same bound for 
$R_{m,\alpha}(t)^{\dag}$.
Thus from \eqref{taratata} we obtain that
\[
\langle D_{\rx}\rangle^{m}\langle\rx\rangle^{\alpha}\p_{t}(W^{\adg}_{\outin}(t)\pi^+ W^{\adg}_{\outin}(t)^{-1})\langle\rx\rangle^{\alpha}\langle D_{\rx}\rangle^{m}\in O(\langle t\rangle ^{-1- \delta + 2 \alpha})
\]
in $B(\cH^{0})$.
This is integrable for $\alpha< \delta/2$.
By integrating from $t=0$ to $t=\pm\infty$,   since $m$ is arbitrary this implies that:
\[
\lim_{t\to\pm\infty}W^{\adg}_{\outin}(t) \pi^+ W^{\adg}_{\outin}(t)^{-1}- \pi^+ \in \langle\rx\rangle^{-\alpha}\cW^{-\infty}(\rr^d)\langle\rx\rangle^{-\alpha}. 
\]
 Since $W^{\adg}_{\outin}= \lim_{t\to\pm\infty}W^{\adg}_{\outin}(t)$ this proves the proposition. \qeds

\subsection{Inhomogeneous Cauchy problem}\label{inhomomo}
Fixing  $\gamma$ with $\12<\gamma<\12 + \delta$, we set:
\[
\cY^{\sobo}\defeq \langle t\rangle^{-\gamma}L^{2}(\rr; H^{\sobo}), \ \ \cY^{\adg, \sobo}\defeq   \langle t\rangle^{-\gamma}L^{2}(\rr; \cH^{\sobo}).
\]
The exponent $\gamma$ is chosen so that  $\cY^{\sobo}\subset L^{1}(\rr; \cH^{\sobo})$, $\cY^{\adg, \sobo}\subset L^{1}(\rr; \cH^{\sobo})$. The benefit of working with $\cY^{(\adg),\sobo}$ rather than with $L^1$ spaces is that the former are Hilbert spaces; this will be needed in Subsect. \ref{ss:fredf}.

\begin{definition}\label{defowit}
 We denote by $\cX^{\sobo}$ the space of $u\in C^{0}(\rr; H^{\sobo+1})\cap C^{1}(\rr; H^{\sobo})$ such that  $Pu\in \cY^{\sobo}$, and similarly by $\cX^{\adg, \sobo}$ the space of $u^{\adg}\in C^{0}(\rr; \cH^{\sobo})$ such that $P^{\adg}u^{\adg}\in \cY^{\adg, \sobo}$.
 We equip $\cX^{(\adg), \sobo}$ with the Hilbert norms:
 \beq\label{defdenorme}\begin{array}{l}
 \|u^{\adg}\|^{2}_{\sobo}\defeq \| \varo^{\adg}_{0}u^{\adg}\|^{2}_{\cH^{\sobo}}+ \| P^{\adg}u^{\adg}\|^{2}_{\cY^{\adg, \sobo}},\\[2mm]
  \|u\|^{2}_{\sobo}\defeq \| \varo_{0}u\|^{2}_{\cE^{\sobo}}+ \| Pu\|^{2}_{\cY^{\sobo}}.
\end{array}
 \eeq
\end{definition}
The existence and uniqueness of the inhomogeneous Cauchy problem for $P$ and $P^{\adg}$ implies  that $\cX^{(\adg), \sobo}$ are Hilbert spaces, as  stated  implicitly in the following lemma.
\begin{lemma}\label{lem:inho} The map
\beq
\bea
\varrho_{0}\oplus P: \cX^{\sobo} &\to  \cE^{\sobo}\oplus\cY^{\sobo}\\
u&\mapsto (\varrho_{0} u, Pu)
\eea
\eeq
is boundedly invertible  with inverse given by:
\beq\label{e12.1}
(\varrho_{0}\oplus P)^{-1}(v, f)= \pi_{0}\cU(t, 0)v-\i\pi_{0} \int_{0}^{t}\cU(t,s)\pi_{1}^{*}f(s)ds.
\eeq
Similarly, the map
\beq
\bea
\varrho^{\adg}_{0}\oplus P^{\adg}: \cX^{\adg, \sobo} &\to  \cH^{\sobo}\oplus\cY^{\adg, \sobo}\\
u^{\adg} &\mapsto (\varrho^{\adg}_{0} u^{\adg}, P^{\adg}u^{\adg})
\eea
\eeq
is boundedly invertible  with inverse given by:
\beq\label{e12.1b}
(\varrho^{\adg}_{0}\oplus P^{\adg})^{-1}(v^{\adg}, f^{\adg})= \cU^{\adg}(t, 0)v^{\adg}+ \i \int_{0}^{t}\cU^{\adg}(t,s)f^{\adg}(s)ds.
\eeq
\end{lemma}
It follows that \begin{equation}
\label{e12.1bb}
\begin{array}{l}
\cX^{\sobo}\hookrightarrow C^{k}(\rr; H^{\sobo+1-k}), \\[2mm]
\cX^{\adg, \sobo}\hookrightarrow C^{k}(\rr; \cH^{\sobo-k}),
\end{array}
\end{equation}
 continuously for $\sobo\in \rr, \ k\in \nn$.\medskip

The following facts are the result of easy computations that make use of \eqref{e100.0}:
\beq\label{e100.1}
\begin{array}{l}
T^{-1}\varrho\in B(\cX^{\sobo}, \cX^{\adg, \sobo+\12}), \ \ -T^{-1}\pi_{1}^{*}\in B(\cY^{\sobo}, \cY^{\adg, \sobo+\12}),\\[2mm]
 \pi_{0}T\in B(\cX^{\adg, \sobo+ \12}, \cX^{\sobo}), \ \ -\pi_{1}T\in B(\cY^{\adg, \sobo+\12}, \cY^{\sobo}).
\end{array}
\eeq
In the sequel we will also need the auxiliary identities\beq\label{e100.2}
\begin{array}{l}
\Ran T^{-1}\varrho= \Ker (\varrho \pi_{0}- \one)T, \\[2mm]
(T^{-1}\varrho)^{-1}= \pi_{0}T\hbox{ on }\Ran T^{-1}\varrho,\\[2mm]
 \Ran T^{-1}\pi_{1}^{*}= \Ker \pi_{0}T,\\[2mm]
 (T^{-1}\pi_{1}^{*})^{-1}=  \pi_{1}T\hbox{ on }\Ran T^{-1}\pi_{1}^{*}.
\end{array}
\eeq

\subsection{Retarded and advanced propagators}\label{retard}
The retarded/advanced propagators for $P^{\adg}$ are defined as follows:
\beq\label{retadv}
(G^{\adg}_{+}f^{\adg})(t)\defeq\i\int_{-\infty}^{t}\cU^{\adg}(t,s )f^{\adg}(s)ds, \ \ (G_{-}^{\adg}f)(t)\defeq -\i \int_{t}^{+\infty}\cU^{\adg}(t, s)f^{\adg}(s)ds,
\eeq
for $f^{\adg}\in L^1(\rr;\cH^{\sobo})$. Using \eqref{e11.51}  one obtains:
\[
\begin{array}{l}
G^{\adg}_{\pm}\in B(L^{1}(\rr; \cH^{\sobo}), C^{0}(\rr; \cH^{\sobo})),\\[2mm]
(G^{\adg}_{\pm})^{\dag}= G^{\adg}_{\mp}\hbox{ on }L^{1}(\rr; \cH^{\sobo}), \ \ P^{\adg}G^{\adg}_{\pm}= \one\hbox{ on }L^{1}(\rr; \cH^{\sobo}).
\end{array}
\]
The analogous propagators for $P$ are:
\begin{equation}
\label{e100.13}
(G_{+}f)(t)=-\i\pi_{0}\int^{t}_{-\infty}\cU(t,s)\pi_{1}^{*}f ds, \ \ (G_{-}f)(t)= \i \pi_{0}\int_{t}^{+\infty}\cU(t, s)\pi_{1}^{*}f(s)ds,
\end{equation}
for $f\in L^{1}(\rr; H^{\sobo})$. One has:
\[
\begin{array}{l}
G_{\pm}\in B(L^{1}(\rr; H^{\sobo}), C^{0}(\rr; H^{\sobo+1})\cap C^{1}(\rr; H^{\sobo})),\\[2mm]
G_{\pm}^{*}= G_{\mp}\hbox{ on }L^{1}(\rr; H^{\sobo}),\ \ PG_{\pm}= \one\hbox{ on }L^{1}(\rr; H^{\sobo}).
\end{array}
\]
Using \eqref{e100.10} the relation between the propagators for $P$ and $P^{\adg}$ is:
\beq\label{e100.13c}
G_{\pm}= - \pi_{0} T G^{\adg}_{\pm}T^{-1}\pi_{1}^{*}.
\eeq
\subsection{Fredholm problems from scattering data}

We now want to define the maps that assign  to an element of $\cX^{(\adg), \sobo}$ its scattering data in the standard sense, as well as its Feynman and anti-Feynman data. By \emph{Feynman data} we mean positive-frequency data of a solution at $+\infty$ and negative-frequency data at $-\infty$, and by \emph{anti-Feynman} the reverse. 

\begin{proposition}\label{idiotic}
 The strong operator limits 
 \[
 \slim_{t\to \pm \infty}\cU_{\outin}^{\adg}(0, t)\varrho^{\adg}_{t}, \ \hbox{resp. }  \slim_{t\to \pm \infty}\cU_{\outin}(0, t)\varrho_{t},
 \]
exist in $B(\cX^{\adg, \sobo}, \cH^{\sobo})$,  resp. in $B(\cX^{\sobo}, \cE^{\sobo})$, 
and equal $(W_{\outin}^{\adg})^{-1}$ on  $\Ker P^{\adg}|_{\cX^{\adg, \sobo}}$, resp. $(W_{\outin})^{-1}$ on $\Ker P|_{\cX^{\sobo}}$.
\end{proposition}
\proof Let $u^{\adg}\in \cX^{\adg, \sobo}$. By Lemma \ref{lem:inho} we have
\[
\bea
\cU_{\out}^{\adg}(0,t)\varrho^{\adg}_{t}u^{\adg}&= \cU_{\out}^{\adg}(0,t)\cU^{\adg}(t, 0)v^{\adg} \\ & \phantom{=} + \i\int_{0}^{t}\cU^{\adg}_{\out}(0, t)\cU^{\adg}(t, 0)\cU^{\adg}(0,s)f^{\adg}(s)ds,
\eea
\]
which by dominated convergence tends to $(W^{\adg}_{\out})^{-1}(v^{\adg}- \varrho^{\adg}_{0}G^{\adg}_{-}f^{\adg})$ as $t\to +\infty$.  Similarly we obtain that $\cU^{\adg}_{\rm in}(0, t)\varrho^{\adg}_{t}u^{\adg}$ converges to $(W_{\rm in}^{\adg})^{-1}(v^{\adg}- \varrho^{\adg}_{0}G^{\adg}_{+}f^{\adg})$ as $t\to -\infty$. 
The proof in the scalar case is similar. 
\qeds
\medskip

We can now introduce  four scattering data maps $\varrho^{(\adg)}_I:\cX^{(\adg), \sobo}\to\cH^{\sobo}$. Note the presence of the operators $T_{\outin}^{-1}$ below; this simplifies some considerations later on.
\begin{definition}
We set:
\[
\begin{array}{l}
\varrho_{\outin }^{\adg}\defeq \slim_{t\to\pm\infty}\cU^{\adg}_{\outin}(0,t)\varrho^{\adg}_{t},\\
\varrho_{\outin }\defeq \slim_{t\to\pm\infty}T_{\outin}^{-1}\cU^{}_{\outin}(0,t)\varrho^{}_{t},\\
\varo^{(\adg)}_{\F}\defeq \pi^{+}\varo^{(\adg)}_{\out}+ \pi^{-}\varo^{(\adg)}_{\inn},\\
\varo^{(\adg)}_{\rm \overline{F}}\defeq \pi^{-}\varo^{(\adg)}_{\out}+ \pi^{+}\varo^{(\adg)}_{\inn}.
\end{array}
\]

\end{definition}

\begin{lemma}\label{l12.1}
 For $I\in\{\inn,\out,\F, \aF\}$ we have:
 \begin{equation}
\label{e100.14}
\varo_{I}= \varo^{\adg}_{I}T^{-1}\varo,
\end{equation}
\beq\label{e100.15}
\bea
\varrho^{\adg}_I &= W_{I}^{\adg\dagger} \circ \varrho^{\adg}_{0} \hbox{ on  } \Ker P^{\adg}|_{\cX^{\adg, \sobo}},\\[2mm]
\varo_{I}&= W_{I}^{\adg\dagger} T^{-1}(0)\varo_{0}\hbox{ on }\Ker P|_{\cX^{\sobo}},
\eea
\eeq
for $I\in\{\inn,\out,\F, \aF\}$, where
\beq\label{eq:defR}
W^{\adg\dagger}_{\F}\defeq  \pi^+W_{\out}^{\adg\dagger}+ \pi^-W_{\inn}^{\adg\dagger}, \quad
W^{\adg\dagger}_{\rm \overline{F}}\defeq  \pi^-W_{\out}^{\adg\dagger}+ \pi^+W_{\inn}^{\adg\dagger}.
\eeq
\end{lemma}
\proof To prove \eqref{e100.14} we write:
\[
\bea
\varo_{\outin}&= T_{\outin}^{-1}\cU_{\outin}(0, t)\varo_{t}+ o(1)
= \cU^{\adg}_{\outin}(0,t)T^{-1}_{\outin}\varo_{t}+ o(1)\\
&=\cU^{\adg}_{\outin}(0,t)T^{-1}(t)\varo_{t}+ o(1)
=\cU^{\adg}_{\outin}(0,t)\varo^{\adg}_{t}T^{-1}\varo+ o(1).
\eea
\]
This implies \eqref{e100.14} for $I= \outin$ and then for $I= \F/{\rm {\overline F}}$.  The first statement of \eqref{e100.15} follows then from the fact that $\varo^{\adg}_{\outin}= W^{\adg\dagger}_{\outin}$ on $\Ker P^{\adg}$,  the second from \eqref{e100.14} and the fact that $T^{-1}\varo: \Ker P\to \Ker P^{\adg}$. \qeds 
 \begin{lemma}\label{l12.2}
 Let $I\in\{ \F, {\rm \overline{F}} \}$. Then 
 $W^{\adg}_{I}  W_{I}^{\adg\dagger}-\one$ and $W_{I}^{\adg\dagger}W^{\adg}_{I} -\one$ are compact  on $\cH^{\sobo}$ and hence  $W^{\adg}_{I}$, $W^{\adg\dag}_{I}$ are Fredholm.  Moreover: 
 \[
 \Ker W^{\adg(\dagger)}_{I}|_{\cH^{\sobo}}= \Ker W^{\adg(\dagger)}_{I}|_{\cE^{\adg,\infty}}, \ \ \coKer W^{\adg(\dagger)}_{I}|_{\cH^{\sobo}}= \coKer W^{\adg(\dagger)}_{I}|_{\cE^{\adg, -\infty}},
  \]
and hence $\ind( W^{\adg(\dag)}_{I})|_{\cH^{\sobo}}$ is independent on $\sobo$.
\end{lemma}
 
\proof We consider  only the $\F$ case. We have
\[
\begin{array}{l}
W^{\adg}_{\F} W^{\adg\dagger}_{\F}=\one+K_{1}, \ \ K_{1} = W^{\adg}_\out \pi ^+ (W^{\adg}_\out)^{-1}-\pi^{+}+W^{\adg}_\inn \pi ^- (W^{\adg}_\inn)^{-1}-\pi^{-},\\[2mm]
W^{\adg\dagger}_{\F} W^{\adg}_{\F}= \one + K_{2}, \ \ K_{2}=\pi^+ (W^{\adg}_{\out})^{-1} W^{\adg}_{\inn} \pi^- + \pi^- (W^{\adg}_{\inn})^{-1}W^{\adg}_\out \pi^+.
\end{array}
\]
By Prop. \ref{prop:stdcase}   we see that $K_{1}$, $K_{2}$ are compact on $\cH^{\sobo}$ and moreover map $\cH^{\sobo}$ to $\cH^{\infty}$. Therefore  $\Ker W_{\F}^{\adg\dag}|_{\cH^{\sobo}}\subset \Ker (\one+ K_{1})|_{\cH^{\sobo}}\subset \cH^{\infty}$ hence $\Ker W_{\F}^{\adg\dag}|_{\cH^{\sobo}}= \Ker W_{\F}^{\adg\dag}|_{\cH^{ \infty}}$. Similarly, identifying $(\cH^{\sobo})^{*}$ with $\cH^{-\sobo}$ and $\coKer A$ with $\Ker A^{*}$ we have $\coKer W_{\F}^{\adg\dag}|_{\cH^{\sobo}}\subset \Ker (\one+ K_{2}^{*})|_{\cH^{\sobo}}\subset \cH^{\infty}$, hence $\coKer W_{\F}^{\adg\dag}|_{\cH^{\sobo}}= \coKer W_{\F}^{\adg\dag}|_{\cH^{-\infty}}$. \qeds



We will need the following lemma, see \cite[Prop. A.1]{BB} for its proof. The next few results are obtained as simple applications of it, following the  strategy  in \cite{BS} (originally applied to the case of the Dirac equation on compact space-times with space-like boundary).

\begin{lemma}\label{lem:fredholm} Let $\cK$ be a Hilbert space and $\cE$, $\cF$ Banach spaces. Let  $K:\cK\to\cE$, $Q:\cK\to\cF$ be bounded and assume that $Q$ is surjective. Then $K:\Ker Q \to \cE$ is Fredholm (of index $l$) iff $K\oplus Q:\cK\to\cE\oplus \cF$ is Fredholm (of index $l$).   
\end{lemma}

\begin{lemma}\label{lem:aux0} For $I\in\{\inn,\out,\F, \aF\}$, the operator
\[
\varrho^{(\adg)}_{I}: \ \{ u^{(\adg)}\in \cX^{(\adg), \sobo}: \ P^{(\adg)}u^{(\adg)}=0\}\to \cH^{\sobo}
\]
is Fredholm of index equal $\ind W_I^{\adg\dag}$ and  is invertible for $I\in\{\inn,\out\}$. 
\end{lemma}
\proof We apply  \eqref{e100.15} and the fact that $\varo_{0}^{\adg}: \Ker P^{\adg}|_{\cX^{\adg, \sobo}}\to \cH^{\sobo}$ and $T^{-1}(0)\varo_{0}: \Ker P|_{\cX^{\sobo}}\to \cH^{\sobo+\12}$ are bijections, by Lemma \ref{lem:inho}. \qeds

\begin{lemma}\label{lem:aux} 
The maps 
\[
\begin{array}{l}
\varrho^{\adg}_{I}\oplus P^{\adg} : \cX^{\adg, \sobo}\to \cH^{\sobo}\oplus \cY^{\adg, \sobo}, \\[2mm]
\varrho_{I}\oplus P : \cX^{\sobo}\to \cH^{\sobo+\12}\oplus \cY^{\sobo},
\end{array}
\] are Fredholm of index $\ind W_{I}^{\adg\dag}$.
\end{lemma}
\proof We use Lemma \ref{lem:fredholm} with $\cK=\cX^{\adg, \sobo}$ resp.  $\cX^{\sobo}$, $\cE=\cH^{\sobo}$ resp. $\cH^{\sobo+\12}$, $\cF=\cY^{\adg, \sobo}$ resp. $\cY^{\sobo}$, $K=\varrho^{\adg}_{I}$ resp. $\varo_{I}$, $Q=P^{\adg}$ resp. $P$. The assumptions of Lemma \ref{lem:fredholm} are satisfied in view of Lemma \ref{lem:aux0} and Lemma \ref{lem:inho} which gives surjectivity of $P^{\adg}$ resp. $P$.  \qeds

Let us introduce the following notation: if $I=\inn/\out$ then $I^{\rm c}\defeq\out/\inn$ and if $I=\F/\aF$ then $I^{\rm c}\defeq \aF/\F$.

\begin{theorem}\label{prop:fredholm1}Assume $(\std)$ with $\delta>1$. Let 
\[
\cX^{(\adg), \sobo}_{I}\defeq\{ u\in \cX^{(\adg), \sobo}: \varrho^{(\adg)}_{I^{\rm c}}u=0\},
\] equipped with the topology of $\cX^{(\adg), \sobo}$.  Then 
\[
\begin{array}{l}
P^{\adg}: \cX^{\adg, \sobo}_{I}\to\cY^{\adg, \sobo},\\[2mm]
P: \cX^{\sobo}_{I}\to \cY^{\sobo}
\end{array}
\] are  Fredholm of index $\ind W_{I^{\rm c}}^{\adg\dag}$.
\end{theorem}
\proof It suffices to  check the assumptions of Lemma \ref{lem:fredholm}  for $\cK=\cX^{\adg, \sobo}$ resp. $\cX^{\sobo}$,  $\cE=\cY^{\adg, \sobo}$ resp. $\cY^{\sobo}$, $\cF=\cH^{\sobo}$ resp. $\cH^{\sobo+\12}$,  $K=P^{\adg}$ resp. $P$, and $Q=\varrho^{\adg}_{I^{\rm c}}$ resp. $\varo_{I^{\rm c}}$. 
 The Fredholm property of $K\oplus Q$ follows from Lemma \ref{lem:aux}, so it remains to check that $\varrho^{\adg}_{I^{\rm c}}:\cX^{\adg, \sobo}\to\cH^{\sobo}$  and $\varo_{I^{\rm c}}: \cX^{\sobo}\to \cH^{\sobo+\12}$ are  surjective.  This is obvious if $I= \outin$ using \eqref{e100.15} and Lemma \ref{lem:inho}.  Let us now consider the case $I= \F$.

Let $\eta_\outin\in\cf(\rr)$ with  $\eta_\inn(t)+\eta_\out(t)=1$ and $\eta_{\outin}(t)=1$ for large $\pm t$. Then 
\[\varrho^{(\adg)}_{\outin}\circ\eta_{\outin}=\varrho^{(\adg)}_{\outin}, \ \varrho^{(\adg)}_{\inout}\circ\eta_{\outin}=0.
\] Furthermore $\eta_\outin \Ker P^{(\adg)}|_{\cX^{(\adg), \sobo}}\subset \cX^{(\adg), \sobo}$. It follows that
\[
\bea
\varrho^{\adg}_{\F} \cX^{\adg, \sobo} &\supset \varrho^{\adg}_{\F}(\eta_\inn\Ker P^{\adg}|_{\cX^{\adg, \sobo}}+\eta_\out \Ker P^{\adg}|_{\cX^{\adg, \sobo}} )\\
&= (\pi^+ \varrho^{\adg}_\out + \pi^-\varrho^{\adg}_\inn )(\eta_\inn\Ker P^{\adg}|_{\cX^{\adg, \sobo}}+\eta_\out \Ker P^{\adg}|_{\cX^{\adg, \sobo}} )\\
&=\pi^+ \varrho^{\adg}_\out  \Ker P^{\adg}|_{\cX^{\adg, \sobo}} + \pi^-\varrho^{\adg}_\inn   \Ker P^{\adg}|_{\cX^{\adg, \sobo}} \\ &= \pi^+\cH^{\sobo} + \pi^- \cH^{\sobo}=\cH^{\sobo}.
\eea
\] 
This proves $\varrho^{\adg}_{\F}:\cX^{\adg, \sobo}\to\cH^{\sobo}$ is surjective.  The same argument shows that $\varo_{\F}: \cX^{\sobo}\to \cH^{\sobo+\12}$ is surjective. In the analogous way we obtain surjectivity of $\varrho^{\adg}_\aF$ and $\varrho_\aF$. \qed

\subsection{Retarded/advanced propagators}
We now show that as anticipated, the retarded/ad\-vanced propagators $G^{\adg}_{\pm}$ are the inverses of $P^{\adg}: \cX^{\adg, \sobo}_{\outin}\to \cY^{\adg, \sobo}$, and a similar statement holds true in the scalar case.

\begin{proposition}\label{proporetard}
 $P^{(\adg)}: \cX^{(\adg), \sobo}_{\outin}\to \cY^{(\adg), \sobo}$ are boundedly invertible with inverse equal to $G^{(\adg)}_{\pm}$.
\end{proposition}
\proof We only treat the case of  $G^{(\adg)}_{+}$, the other one being analogous. We have seen in Subsect. \ref{retard} that 
\[
G^{\adg}_{+}\in B(L^{1}(\rr; \cH^{\sobo}),C^{0}(\rr; \cH^{\sobo}))
\]
and  $P^{\adg} G^{\adg}_{+}=\one$ on $L^{1}(\rr; \cH^{\sobo})$,  hence $G^{\adg}_{+}\in B(\cY^{\adg, \sobo}, \cX^{\adg, \sobo})$ and $P^{\adg}G^{\adg}_{+}=\one$ on $\cY^{\adg, \sobo}$. Since $\lim_{t\to-\infty}\varrho^{\adg}_{t} G^{\adg}_{+}f^{\adg}=0$,  we have $G^{\adg}_{+}\cY^{\adg, \sobo}\subset \cX^{\adg, \sobo}_{\out}$. It remains to show that $G^{\adg}_{+}P^{\adg}=\one$ on $\cX^{\adg, \sobo}_{\out}$. If $u^{\adg}\in \cX^{\adg, \sobo}_{\out}$  we have:
\[
\bea
(G^{\adg}_{+}P^{\adg}u^{\adg})(t)&= \int_{-\infty}^{t}\cU^{\adg}(t, s)(\pe_{s}- \i H^{\adg}(s))u^{\adg}ds\\
&=\lim_{\varT\to -\infty}\int_{-\infty}^{t}\cU^{\adg}(t, s)(\pe_{s}- \i H^{\adg}(s))u^{\adg}ds\\
&=\lim_{\varT\to -\infty}\left[\cU^{\adg}(t,s)u(s)\right]^{t}_{\varT}\\&\phantom{\,=}-\lim_{\varT\to -\infty}\int_{\varT}^{t}(-\pe_{s}+ \i H^{\adg}(s))\cU^{\adg}(t,s)u^{\adg}(s)ds\\
&=u^{\adg}(t),
\eea
\]
since $\lim_{\varT\to -\infty}u^{\adg}(\varT)=0$ in view of $u^{\adg}\in \cX^{\adg, \sobo}_{\out}$. 
In the scalar case we obtain from \eqref{e100.0} that $(D_{t}- \AH(t))TG^{\adg}_{+}T^{-1}=\one$ hence $(\varo \pi_{0}-\one)TG^{\adg}_{+}T^{-1}\pi_{1}^{*}=0$ which implies that $PG_{+}=\one$ on $\cY^{\sobo}$.  Conversely, by \eqref{e100.1}, \eqref{e100.14} we know that $T^{-1}\varo: \cX^{\sobo}_{\out}\to \cX^{\adg, \sobo+\12}_{\out}$. Since $G^{\adg}_{+}P^{\adg}= \one$ on $\cX^{\adg, \sobo+ \12}$ this yields
\[
TG^{\adg}_{+}T^{-1}(D_{t}- \AH(t))\varo=TG^{\adg}_{+}P^{\adg}T^{-1}\varo= \varo\hbox{ on }\cX^{\sobo}_{\out},  
\]
hence $G_{+}P= \one$ on $\cX^{\sobo}_{\out}$ using  $(D_{t}-\AH(t))\varo= \pi_{1}^{*}\pi_{1}(D_{t}- \AH(t))$. This completes the proof. \qeds
\subsection{ The Fredholm inverses for $P^{(\adg)}$ on $\cX^{(\adg), \sobo}_{\F}$}
From Thm.  \ref{prop:fredholm1} we know that $P^{(\adg)}: \cX^{(\adg), \sobo}_{\F}\to \cY^{(\adg), \sobo}$ are Fredholm. We will now construct explicit approximate inverses $G^{(\adg)}_{\F}$ of $P^{(\adg)}:\cX^{(\adg),\sobo}_{\F}\to\cY^{\sobo}$, which requires some special care because of the requirement $\varrho_\aF^{(\adg)} \circ    G^{(\adg)}_{\F}=0$ that follows from the definition of $\cX^{(\adg),\sobo}_{\F}$ (in fact, for instance the time-ordered Feynman propagators associated to the \emph{in} or \emph{out} state, see \cite{isozaki,inout}, fail in general to satisfy this condition). We will then show that $G_{\F}$ has the required wave front set.

\subsubsection{Auxiliary diagonal Hamiltonian}\label{auxili} Recall that 
\begin{equation}
\label{e11.5b}
H^{\adg}(t)= H^{\dg}(t)-V^{\adg}_{-\infty}(t),
\end{equation}
Where $V^{\adg}_{-\infty}(t)\in\Psi^{-\infty, -1- \delta}_{\std}(\rr; \rr^{d})\otimes \cc^{2}$ and $H^{\dg}(t)$ is the diagonal matrix  with on-diagonal components $\epsilon^{\pm}(t)$.

Let $\cU^{\dg}(t,s)$ be the evolution generated by  $H^{\dg}(t)$. Using \eqref{e11.5b}  we see that $\cU^{\dg}(t,s)$ is well defined  and moreover, 
\[
\sup_{t, s\in \rr}\|\cU^{\dg}(t,s)\|_{B(\cH^{\sobo})}<\infty,
\]
using the same argument as for $\cU(t,s)$, see \cite{inout}.  Since $H^{\dg}(t)= H^{\dg\dag}(t)$ we also have
\beq\label{e11.5c}
\cU^{\dg}(t,s)^{\dag}= \cU^{\dg}(s,t).
\eeq
We set correspondingly
\[
P^{\dg}\defeq D_{t}- H^{\dg}(t)= P^{\adg} - V^{\adg}_{-\infty}(t).
\]
Note that since $\| V^{\adg}_{-\infty}(t)\|_{B(\cH^{\sobo})}= O(\langle t\rangle^{-1- \delta})$ and  we have assumed that $\gamma<\12 + \delta$, we see that for $u\in C^{0}(\rr; \cH^{\sobo})$ we have $ P^{\adg}u\in \cY^{\adg, \sobo}$ if and only if $P^{\dg}u\in \cY^{\adg, \sobo}$ and the two norms in \eqref{defdenorme} on $\cX^{\adg, \sobo}$ defined with $P^{\adg}$ and $P^{\dg}$ are equivalent.

\subsubsection{Fredholm inverse for $P^{\adg}$ on $\cX^{\adg, \sobo}_{\F}$}
\begin{definition}We set for $f^{\adg}\in \cY^{\adg, \sobo}$:
 \[
\bea
G^{\adg}_{\F} f^{\adg}(t)&\defeq  \i\int_{-\infty}^{t}\cU^{\dg}(t, 0)\pi^{+}\cU^{\dg}(0,s)f^{\adg}(s)ds\\ &\phantom{=\,}-\i\int_{t}^{+\infty}\cU^{\dg}(t, 0)\pi^{-}\cU^{\dg}(0,s)f^{\adg}(s)ds.
\eea
\]
\end{definition}
Using the `time-kernel notation' $A(t,s)\defeq  \varrho^{\rm ad}_{t}\circ A\circ (\varrho^{\rm ad}_{s})^{*}$ we can write:
\[
\bea
G^{\adg}_{\F}(t, s)&= \i \theta(t-s)\cU^{\dg}(t, 0)\pi^{+}\cU^{\dg}(0,s)-\i \theta(s-t)\cU^{\dg}(t, 0)\pi^{-}\cU^{\dg}(0,s)\\
&= \i \cU^{\dg}(t, 0)\pi^{+}\cU^{\dg}(0,s)-\i \theta(s-t)\cU^{\dg}(t,s),
\eea
\]
where $\theta$ is the Heaviside step function. Let us also observe that since $[\cU^{\dg}(t,s), \pi^{+}]=0$, we have
\beq\label{asz}
G^{\adg}_{\F}= G^{\dg}_{+}\pi^{+}+ G^{\dg}_{-}\pi^{-},
\eeq
where $G^{\dg}_{\pm}$ are the retarded/advanced propagators for $H^{\dg}(t)$, defined in analogy to $G^{\adg}_{\pm}$.

\begin{theorem}\label{theotheo}
Let $m\in\rr$. We have:
\[
\begin{array}{rl}
i)&G^{\adg}_{\F}\in B(\cY^{\adg, \sobo}, \cX^{\adg, \sobo}_{\F}), \  P^{\adg}G^{\adg}_{\F}= \one_{\cY^{\adg, \sobo}}+ K_{\cY^{\adg, \sobo}}, \\[1mm] &\hbox{where }K_{\cY^{\adg, \sobo}}\hbox{ is compact on }\cY^{\adg, \sobo} \\[2mm]
ii)&G^{\adg}_{\F} P^{\adg}= \one_{\cX^{\adg, \sobo}_{\F}}+ K_{\cX^{\adg, \sobo}_{\F}}, \hbox{where }K_{\cX^{\adg, \sobo}_{\F}}\hbox{ is compact on }\cX^{\adg, \sobo}_{\F},\\[2mm]
iii)&\i^{-1}q^{\adg}(G^{\adg}_{\F}- (G^{\adg}_{\F})^{\dag})\geq 0\hbox{ on }\cY^{\adg, \sobo}, \hbox{ for }\sobo\geq 0.
\end{array}
\] 
\end{theorem}
To prove Thm. \ref{theotheo} we will need the following lemma.
\begin{lemma}\label{compact1}
$V^{\adg}_{-\infty}: \cX^{\adg, \sobo}\to \cY^{\adg, \sobo}$ is compact.
\end{lemma}
\proof 
From \eqref{e12.1bb} we first obtain that the injection $\cX^{\adg, \sobo}\hookrightarrow C^{k+1}(\rr; \cH^{\sobo-k-1})$ is bounded for any $k\in \nn$, $\sobo\in \rr$. We pick $\varepsilon>0$ such that $\gamma<\12 + \delta -2\varepsilon$ and write $V^{\adg}_{-\infty}(t)$ as $\langle t\rangle^{-1- \delta+\varepsilon}\langle \rx\rangle^{-\varepsilon}Y^{\adg}(t)$, where 
$Y^{\adg}(t)\in C^{\infty}(\rr; \cW^{-\infty}(\rr^{d})\otimes B(\cc^2))$. It follows that $Y^{\adg}: C^{k}(\rr; \cH^{\sobo})\to C^{k}(\rr; \cH^{\sobo'})$ is bounded for any $\sobo,\sobo'$, hence 
\[
V^{\adg}_{-\infty}: \cX^{\sobo}\to \langle t\rangle^{-1- \delta + \varepsilon}C^{k+1}(\rr; \langle \rx\rangle^{-\varepsilon}\cH^{\sobo'+1})\hbox{ is bounded for any } k\in \nn, \ \sobo, \sobo'\in \rr. 
\]
This implies that 
\beq\label{compacto}
V^{\adg}_{-\infty}: \cX^{\sobo}\to \langle t\rangle^{-1- \delta + 2\varepsilon}C^{k}(\rr; \cH^{\sobo'})\hbox{ is compact for any }k\in \nn, \ \sobo, \sobo'\in \rr. 
\eeq
We use \eqref{compacto} for   $k=0,m'=m$,   and the fact that  the injection  $\langle t\rangle^{-1- \delta + 2\varepsilon}C^{0}(\rr; \cH^{\sobo})\hookrightarrow \langle t\rangle^{-\gamma}L^{2}(\rr; \cH^{\sobo})= \cY^{\adg, \sobo}$ is bounded since $\gamma<\12+ \delta-2 \varepsilon$. It follows that $V^{\adg}_{-\infty}: \cX^{\adg, \sobo}\to \cY^{\adg, \sobo}$ is compact.  \qeds

{\noindent\bf Proof of Thm. \ref{theotheo}.}
{\it Proof of $i)$}: note first that  $G^{\adg}_{\F}= \gdia_{+}\pi^{+}+ \gdia_{-}\pi^{-}\in B(\cY^{\adg, \sobo}, \cX^{\adg, \sobo})$ since $\gdia_{\pm}\in B(\cY^{\adg, \sobo}, \cX^{\adg, \sobo})$.
We then have:
\[
\bea
P^{\adg} G^{\adg}_{\F}&= P^{\dg}G^{\adg}_{\F}+ V^{\adg}_{-\infty} G^{\adg}_{\F}\\[2mm]
&= P^{\dg}G^{\dg}_{+}\pi^{+}+ P^{\dg}G^{\dg}_{-}\pi^{-}+ V^{\adg}_{-\infty} G^{\adg}_{\F}\\[2mm]
&=\one_{\cY^{\adg, \sobo}}+ V^{\adg}_{-\infty} G^{\adg}_{\F},
\eea
\]
by Prop. \ref{proporetard} applied to $P^{\dg}$.  By Lemma \ref{compact1}, $V^{\adg}_{-\infty}G^{\adg}_{\F}$ is compact on $\cY^{\adg, \sobo}$.

 It remains to check that 
$G^{\adg}_{\F}: \cY^{\adg, \sobo}\to\cX^{\adg, \sobo}_{\F}$, i.e.  $\pi^{+}\varrho^{\adg}_{\rm in}G^{\adg}_{\F}= \pi^{-}\varrho^{\adg}_{\out}G^{\adg}_{\F}=0$. We have:
 \[
 \pi^{+}\varrho^{\adg}_{\rm in}G^{\adg}_{\F}= \varrho^{\adg}_{\rm in}\pi^{+}(\gdia_{+}\pi^{+}+ \gdia_{-}\pi^{-})= \varrho^{\adg}_{\rm in}\gdia_{+}\pi^{+}=0,
 \]
 since $[\gdia_{\pm}, \pi^{\pm}]=0$ and $\varrho^{\adg}_{\rm in }\gdia_{+}=0$. Similarly we obtain that $ \pi^{-}\varrho^{\adg}_{\out}G^{\adg}_{\F}=0$,  which completes the proof of {\it i)}.
 
 {\it Proof of $ii)$}:   we have  by \eqref{asz}:
 \[
G^{\adg}_{\F} P^{\adg}= G^{\adg}_{\F} P^{\dg}+ G^{\adg}_{\F} V^{\adg}_{-\infty}=\gdia_{+}P^{\dg}\pi^{+}+ \gdia_{-}P^{\dg}\pi^{-}+ G^{\adg}_{\F} V^{\adg}_{-\infty}.
\]
 If $u\in \cX^{\adg, \sobo}_{\F}$ we have $\pi^{+}\varrho^{\adg}_{\rm in }u= \varrho^{\adg}_{\rm in }\pi^{+}u=0$ and  $\pi^{-}\varrho^{\adg}_{\rm out }u= \varrho^{\adg}_{\rm out }\pi^{-}u=0$. By Prop. \ref{proporetard} applied to $P^{\dg}$ we have $\gdia_{+}P^{\dg}\pi^{+}u= \pi^{+}u$, $\gdia_{-}P^{\dg}\pi^{-}u= \pi^{-}u$, hence
 \[
G^{\adg}_{\F} P^{\adg}= \one_{\cX^{\adg, \sobo}_{\F}}+ G^{\adg}_{\F}V^{\adg}_{-\infty}.
\]
Again, by Lemma \ref{compact1} $G^{\adg}_{\F}V^{\adg}_{-\infty}$ is compact on $\cX^{\adg, \sobo}$. 
 
{\it Proof of iii)}: using the time-kernel notation we have by \eqref{e11.5c}
\[
\bea
&(G^{\adg}_{\F})^{\dag}(t,s)= G^{\adg}_{\F}(s,t)^{\dag}\\[2mm]
&=  \i \theta(t-s)\cU^{\dg}(t,0)\pi^{-}\cU^{\dg}(0,s)- \i \theta(s-t)\cU^{\dg}(t,0)\pi^{+}\cU^{\dg}(0,s).
\eea
\]
Hence, defining  $\cU^{\dg}\in B(\cH^{\sobo}, \cX^{\adg, \sobo})$ by 
  \beq\label{defudiag}
	\bea 
  \cU^{\dg}v^{\adg}(t)\defeq  \cU^{\dg}(t,0)v^{\adg}, \ \ v^{\adg}\in \cH^{\sobo},
  \eea \eeq 
we have
\[
\i^{-1}(G^{\adg}_{\F}- (G^{\adg}_{\F})^{\dag})(t,s)= \cU^{\dg}(t,0)(\pi^{+}- \pi^{-})\cU^{\dg}(0,s)=\cU^{\dg}(t,0)q^{\adg}\cU^{\dg}(0,s).
\]
It follows that
\[
\i^{-1}(f^{\adg}| q^{\adg}(G^{\adg}_{\F}- (G^{\adg}_{\F})^{\dag})f^{\adg})_{\cH^{0}}= ((\cU^{\dg})^{\dag}f^{\adg}| (q^{\adg})^{2}(\cU^{\dg})^{\dag}f^{\adg})_{\cH^{0}}\geq 0,
\]
hence $\i^{-1}q^{\adg}(G^{\adg}_{\F}- (G^{\adg}_{\F})^{\dag})\geq 0$ on $\cH^{0}$  hence on $\cH^{\sobo}$ for $\sobo\geq 0$. \qeds

\subsubsection{Fredholm inverse for $P$ on $\cX^{\sobo}_{\F}$}
\begin{theorem}\label{teuheuteuheu} Assume $(\std)$ with $\delta>1$. Let \beq\label{e100.20}
G_{\F}\defeq- \pi_{0} T G^{\adg}_{\F}T^{-1}\pi_{1}^{*}.
\eeq
We have:
\[
\begin{array}{rl}
i)&G_{\F}\in B(\cY^{\sobo}, \cX^{\sobo}_{\F}), \  PG_{\F}= \one_{\cY^{\sobo}}+ K_{\cY^{\sobo}}, \hbox{where }K_{\cY^{\sobo}}\hbox{ is compact on }\cY^{\sobo},\\[2mm]
ii)&G_{\F}P= \one_{\cX^{\sobo}_{\F}}+ K_{\cX^{\sobo}_{\F}}, \hbox{where }K_{\cX^{\sobo}_{\F}}\hbox{ is compact on }\cX^{\sobo}_{\F},\\[2mm]
iii)&\i^{-1}(G_{\F}- G_{\F}^{*})\geq 0\hbox{ on }\cY^{\sobo}, \hbox{ for }\sobo\geq 0,\\[2mm]
iv)&P G_{\F}-\one, \ G_{\F}P-\one \hbox{ are smoothing operators},\\[2mm]
v)& \WF(G_{\F})'= (\diag_{T^*M})\cup\textstyle\bigcup_{t\leq 0}(\Phi_t(\diag_{T^*M})\cap \pi^{-1}\cN).
\end{array}
\] 
In particular $G_{\F}$ is a Feynman parametrix of $P$ in the sense of Def. \ref{def:Fp}.
\end{theorem}
\proof 
{\it Proof of $i)$:} from \eqref{e100.1} and Thm. \ref{theotheo} we see that $G_{\F}\in B(\cY^{\sobo}, \cX^{\sobo})$. Let us show that $G_{\F}$ maps $\cY^{\sobo}$ into $\cX^{\sobo}_{\F}$.   For $V^{\adg}_{-\infty}$ the operator introduced in \eqref{e11.5b} we have:
\[
P^{\adg} G^{\adg}_{\F}= \one +  V^{\adg}_{-\infty}G^{\adg}_{\F} \ \Rightarrow \ TP^{\adg}T^{-1}TG^{\adg}_{\F}T^{-1}= \one + TV^{\adg}_{-\infty} G^{\adg}_{\F}T^{-1}.
\]
Using \eqref{e100.0} this implies that:
\beq\label{e100.4}
(D_{t}- \AH(t))T G^{\adg}_{\F}T^{-1}\pi_{1}^{*}= \pi_{1}^{*}+ TV^{\adg}_{-\infty}G^{\adg}_{\F}T^{-1}\pi_{1}^{*}\eqdef \pi_{1}^{*}+ R_1,
\eeq
where $R_1\in B(\cY^{\sobo}, \langle t\rangle^{-1-\delta}C^{0}(\rr; \cE^{\sobo}))$, using that $V^{\adg}_{-\infty}\in \Psi^{-\infty, -1- \delta}_{\std}(\rr; \rr^{d})\otimes B(\cc^2)$.
This implies that
\[
\varrho \pi_{0} T G^{\adg}_{\F}T^{-1}\pi_{1}^{*}= T G^{\adg}_{\F}T^{-1}\pi_{1}^{*}+ R_2,
\]
where $R_2\in B(\cY^{\sobo}, \langle t\rangle^{-1- \delta}C^{0}(\rr; \cE^{\sobo}))$. We now have:
\[
\bea
\cU_{\outin}(0,t)\varrho_{t}G_{\F}f&=-\cU_{\outin}(0,t)\varrho_{t}\pi_{0} T G^{\adg}_{\F}T^{-1}\pi_{1}^{*}f\\[2mm]
&=-\cU_{\outin}(0,t)T(t)\varrho^{\adg}_{t} G^{\adg}_{\F}T^{-1}\pi_{1}^{*}f+ o(1)\\[2mm]
&=-T_{\outin}\cU^{\adg}_{\outin}(0,t)T_{\outin}^{-1}T(t)\varrho^{\adg}_{t}G^{\adg}_{\F}T^{-1}\pi_{1}^{*}f+ o(1)\\[2mm]
&=-T_{\outin}\cU^{\adg}_{\outin}(0,t)\varrho^{\adg}_{t}G^{\adg}_{\F}T^{-1}\pi_{1}^{*}f+ o(1)\\[2mm]
&=- T_{\outin}\varrho^{\adg}_{\outin} G^{\adg}_{\F}T^{-1}\pi_{1}^{*}f+ o(1),
\eea
\]
hence
\begin{equation}
\label{e100.3}
\varrho_{\outin}G_{\F}= - \varrho^{\adg}_{\outin} G^{\adg}_{\F}T^{-1}\pi_{1}^{*}.
\end{equation}
By Thm. \ref{theotheo} we have $\varrho^{\adg}_{\aF}G^{\adg}_{\F}=0$, i.e. $\pi^{-}\varrho^{\adg}_{\out}G^{\adg}_{\F}= \pi^{+}\varrho^{\adg}_{\inn}G^{\adg}_{\F}=0$, which by \eqref{e100.3} gives $\varrho_{\aF}G_{\F}=0$.
It follows that $G_{\F}$ maps $\cY^{\sobo}$ to $\cX^{\sobo}_{\F}$ as claimed.

From \eqref{e100.4}, we obtain by an easy computation:
\beq\label{e100.5}
P G_{\F}= \one - \pi_{1}R_1- D_{t}\pi_{0}R_1+ \i r\pi_{0}R_1.
\eeq
Using \eqref{compacto} we obtain that $R_1: \cY^{\sobo}\to \langle t\rangle^{-1- \delta + 2\varepsilon}C^{k}(\rr;  \cE^{\sobo'})$ is compact for any $\sobo, \sobo', k$, hence $PG_{\F}- \one$ is compact on $\cY^{\sobo}$. 

{\it Proof of $ii)$:} by Thm. \ref{theotheo} and \eqref{e100.0} we know that:
\[
G^{\adg}_{\F}T^{-1}(D_{t}- \AH(t))T= G^{\adg}_{\F}P^{\adg}= \one + G^{\adg}_{\F}V^{\adg}_{-\infty}\hbox{ on }\cX^{\adg, \sobo+ \12}_{\F},
\]
hence
\[
TG^{\adg}_{\F}T^{-1}(D_{t}- \AH(t))T= T+ TG^{\adg}_{\F}V^{\adg}_{-\infty}\hbox{ on }\cX^{\adg, \sobo+ \12}_{\F}.
\]
By \eqref{e100.1}, \eqref{e100.14} we know that $T^{-1}\varrho: \cX^{\sobo}_{\F}\to \cX^{\adg, \sobo+ \12}_{\F}$. It follows that
\[
TG^{\adg}_{\F}T^{-1}(D_{t}- \AH(t))\varrho= \varrho + T G^{\adg}_{\F}V^{\adg}_{-\infty}T^{-1}\varrho, \hbox{ on }\cX^{\sobo}_{\F}.
\]
Since $(D_{t}-\AH(t))\varo= \pi_{1}^{*}\pi_{1}(D_{t}- \AH(t))$, we obtain that
\beq\label{e100.5b}
\bea
G_{\F}P&= \pi_{0}TG^{\adg}_{\F}T^{-1}\pi_{1}^{*}\pi_{1}(D_{t}- \AH(t))\varo\\[2mm]
&=\pi_{0}\varo+   \pi_{0}T G^{\adg}_{\F}V^{\adg}_{-\infty}T^{-1}\varo\\[2mm]
&=\one +  \pi_{0}T G^{\adg}_{\F}V^{\adg}_{-\infty}T^{-1}\varo  \hbox{ on }\cX^{\sobo}_{\F}.
\eea
\eeq
Using \eqref{e100.1} and Lemma \ref{compact1} we obtain that $G_{\F}P-\one$ is compact on $\cX^{\sobo}_{\F}$, which proves {\it ii)}.

{\it Proof of $iii)$:} we note that for any operator $A^{\adg}$ one has  
\[
\pi_{0}A^{\adg}\pi_{1}^{*}= \pi_{1}q A^{\adg}\pi_{1}^{*}, \ \hbox{ hence } (\pi_{0}A^{\adg}\pi_{1}^{*})^{*}= \pi_{1}q A^{\adg\dag}\pi_{1}^{*}.
\]
This gives
\[
\i(G_{\F}- G_{\F}^{*})= \i^{-1}\pi_{1}q T(G^{\adg}_{\F}- G^{\adg\dag}_{\F})T^{-1}\pi_{1}^{*}=\i^{-1}(\pi_{1}T)q^{\adg}(G^{\adg}_{\F}- G^{\adg\dag}_{\F})(\pi_{1}T)^{*}\geq 0,
\]
by Thm. \ref{theotheo} {\it iii)}.

{\it Proof of $iv)$:} we first  see using $\cU^{\dg}(t,s)(D_{s}- H^{\dg}(s))=0$ and integration by parts that  $G^{\adg}_{\F}$ maps compactly supported elements of $H^{-p}(\rr; \cH^{-k})$ into $H^{p}(\rr; \cH^{-k-2p})$ for $k, p\in \nn$, hence $V^{\adg}_{-\infty}G^{\adg}_{\F}$ maps compactly supported elements of $H^{-p}(\rr; \cH^{k})$ into $H^{p}(\rr, \cH^{k})$. The same argument shows that $G^{\adg}_{\F}$ maps also compactly supported elements $H^{-p}(\rr; \cH^{k})$ into $H^{p}(\rr, \cH^{k})$. By \eqref{e100.4}, \eqref{e100.5}, \eqref{e100.5b} this implies {\it iv)}.

{\it Proof of $v)$:} Let $G_{\F, \rf}$ be given by the analog of \eqref{e100.20} with $\cU^{\dg}$ replaced by $\cU^{\adg}$ in the definition of $G^{\adg}_{\F}$. By  \cite[Thm. 7.10, Prop. 7.11]{bounded}, $G_{\F, \rf}$ is a Feynman parametrix. From \eqref{turlututu} we obtain that $\cU^{\dg}(t,s)- \cU^{\adg}(t,s)\in \cinf(\rr^{2}; \cW^{-\infty}(\rr^{d}))$, which implies that $G_{\F}- G_{\rm F, \rf}$ is smoothing and completes the proof of of {\it v)}. \qed

\section{Asymptotically Minkowski spacetimes}\init\label{sec:ams}

\subsection{Assumptions}\label{ss:nt}
In this final section we consider asymptotically Minkowski spacetimes and prove analogues of the results from Sect. \ref{sec:abstract} by a reduction procedure.

We work on $M= \rr^{1+d}$ and use the notation $y=(t, \ry)$ for its elements.

For $\delta\in \rr$ we denote by $S_{\std}^{\delta}(\rr^{1+d})$, resp. $S_{\std}^{\delta}(\rr^\pm\times\rr^{d})$, the class of smooth functions such that 
\[
\p^{\alpha}_{y}f\in O(\langle y\rangle^{\delta- |\alpha|}), \ \alpha\in \nn^{1+d},
\]
holds on $\rr^{1+d}$, resp. $\rr^\pm \times \rr^d$. 

The analogous spaces on $\rr^{d}$ will be denoted by $S^{\delta}_{\rm sd}(\rr^{d})$.

We denote by $\alteta_{\mu\nu}$ the Minkowski metric on $\rr^{1+d}$, fix a Lorentzian metric $\altg$ on $\rr^{1+d}$ and  consider the Klein-Gordon operator
\begin{equation}
\label{e11.1}
P= -\Box_{\altg}+ \altV\,(y),
\end{equation}
where  $\altV$ is again a smooth real function on $\rr^{1+d}$.  We assume that $(M, \altg)$ is asymptotically Minkowski and $\altV$ is asymptotically constant in the following sense:
\[
(\aM)\ \begin{array}{rl}
i)&\altg_{\mu\nu}(y)- \alteta_{\mu\nu} \in S^{-\delta}_{\std}(\rr^{1+d}), \ \delta>1,\\[2mm]
ii)&\altV(y)- \altm^{2}\in S^{-\delta}_{\std}(\rr^{1+d}), \ \altm>0,\ \delta>1, \\[2mm]
iii)&(\rr^{1+d}, \altg) \hbox{ is globally hyperbolic},\\[2mm]
iv)&(\rr^{1+d}, \altg) \hbox{ has a  time function }\tilde{t}\hbox{ with  }\tilde{t}- t\in S^{1-\epsilon}_{\std}(\rr^{1+d})\hbox{ for }\epsilon>0.
\end{array}
\]
We recall that a smooth function $\tilde{t}$ is called a  \emph{time function} if  $\nabla \tilde{t}$ is a timelike vector field.

\begin{remark}
 We conjecture that $(\aM) \ iv)$  follows from $(\aM)\ i), \ iii)$.
\end{remark}
\subsection{Global hyperbolicity and non-trapping condition} In what follows we discuss the relation of our hypothesis $(\aM)$ with non-trapping assumptions for null geodesics.

The null geodesics of $\altg$ coincide modulo reparametrization with  the projections on the base of  null bicharacteristics of 
$m(x, \xi)= |\xi|^{-1}\xi\cdot \altg^{-1}(x)\xi$.  We recall that 
$\Phi_{s}$  is  the flow of the Hamiltonian vector field $H_{p}$, $p(y, \xi)=\xi\cdot \altg^{-1}(y)\xi$, which acts naturally on $S^{*}\rr^{1+d}= T^{*}\rr^{1+d}\cap \{|\xi|=1\}$.  Null bicharacteristics stay in one of the two connected components $\cN^{\pm}$ of $\cN$. We set
\[
\Gamma_{\rm in /out}^{\pm}=\{X\in S^{\pm}: \ \phi_{s}(X)\not\to \infty \hbox{ as }s\to \pm \infty \}, \ S^{\pm}= \cN^{\pm}\cap \{|\xi|=1\}.
\]
The familiar \emph{non-trapping condition} is:
\[
 (\nt) \ \hbox{ there are no trapped null geodesics of }\altg, \hbox{ i.e. } \Gamma^{\pm}= \Gamma^{\pm}_{\rm in}\cap \Gamma^{\pm}_{\rm out}= \emptyset.
 \]
 By a well-known argument, this actually implies that $\Gamma^{\pm}_{\rm in/out}= \emptyset$, see Lemma \ref{lem:traop} below, hence any null geodesic escapes to infinity  both when the affine parameter $s$ tends to $+\infty$ {\em and} to $-\infty$. 

\begin{lemma}\label{lem:traop}If $(\nt)$ holds then $\Gamma^{\pm}_{\rm in}= \Gamma^{\pm}_{\rm out}=\emptyset$.
\end{lemma}
\proof We  drop the $\pm$ superscript. We claim that $\Gamma_{\rm in}\neq \emptyset$  or $\Gamma_{\rm out}\neq \emptyset$ implies $\Gamma\neq \emptyset$. In fact Let $X_{0}\in \Gamma_{\rm in}$, $K_{1}$ a compact set such that $\{\Phi_{s}(X_{0}) : \ s\geq 0\}\subset K_{1}$. Let $s_{j}\to +\infty$ a sequence such that $X_{j}= \Phi_{s_{j}}(X_{0})\to X_{\infty}\in K_{1}$. Clearly $\Phi_{s}(X_{j})= \Phi_{s+s_{j}}(X_{0})\to \Phi_{s}(X_{\infty})$ for any $s\in \rr$. For $j$ large enough we have $\Phi_{s+s_{j}}(X_{0})\in K_{1}$ hence $\Phi_{s}(X_{\infty})\in K_{1}$ for any $s\in \rr$, which means that $X_{\infty}\in \Gamma$.
\qeds
\begin{proposition}\label{propnico}
 Assume $(\aM)\ i)$. Then 
 \ben
 \item $(\rr^{1+d}, \altg)$ is globally hyperbolic iff $(\nt)$ holds,
 \item if $(\aM)$, $iii)$ and $iv)$ hold then there exists a Cauchy time function $\tilde{t}$ such that $\tilde{t}- t\in \coinf(\rr^{1+d})$.
 \een
\end{proposition}
In the sequel we will work with the Cauchy time function $\tilde{t}$ obtained in  (2) of Prop. \ref{propnico} .\medskip

\proof First let us prove (1). By $(\aM)\  i)$ we have
\[
\{p, t\}=  \p_{\tau}(|\xi|^{-1}(\tau^{2}- k^{2})) + O(\langle x\rangle^{-\delta}|\xi|^{-1})\geq \tau|\xi|^{-1}+ O(\langle x\rangle^{-\delta}|\xi|^{-1}).
\]
It follows that there exist $c_{0}>0$ and compact sets $K^{\pm}\subset \cN^{\pm}$ such that 
\begin{equation}
\label{enice.1}
\pm\{p, t\}\geq c_{0} \hbox{ on }S^{\pm}\setminus K^{\pm}
\end{equation}
This implies that if $X\in S^{+}$ and $\phi_{s}(X)\to \infty$ when $s\to \pm \infty$ then $t\circ \phi_{s}(X)\to \pm\infty$ when $s\to \pm \infty$. Of course a similar statement  is true for $X\in S^{-}$ with  the reversed sign.

Let us set $\Sigma_{s}= t^{-1}(s)$.
Using   \eqref{enice.1} we obtain that there exists $T_{0}>0$ such that any null geodesic intersects $\Sigma_{\pm T}$ transversally for $T\geq T_{0}$ and hence enters $I^{\pm}(\Sigma_{\pm T})$.  Moreover  $\Sigma_{\pm T}$ is achronal for $T$ large enough, since $\p_{t}$ is a future directed time-like vector field in $\{\pm t\geq \pm T\}$ for $T$ large enough.  We can apply then the Geroch-S\'anchez  theorem  (see for instance \cite[Thm. 8.3.7]{W} for its basic version), which implies that $\Sigma_{\pm T}$ are Cauchy hypersurfaces for $T$ large enough, which completes the proof of (1), $\Leftarrow$.

 Assume now  that $(\rr^{1+d}, \altg)$ is globally hyperbolic  and $(\nt)$ is violated. Let  $\gamma=\{x(s) : \ s\in \rr\}$ be  an (affine parametrized) null geodesic  which is past and future trapped, ie $\gamma\subset K$ for some compact  set $K$.  Since $(\rr^{1+d}, \altg)$ is  strongly causal, for each $x\in K$ there exists an open neighborhood $U(x)$ of $x$ such that $\gamma$ enters $U(x)$ only once, ie $\{s\in \rr : \ x(s)\in U(x)\}\defeq I(x)$ is a bounded  open interval.  By compactness  of $K$ we have $x(s)\not\in K$ for $s\not\in \cup_{i=1}^{n}I(x_{i})$, which is a contradiction. This completes the proof of (1), $\Rightarrow$.
 
 Now let us prove  (2).
 Let $\tilde{t}$ be the  time function in $(\aM)\ iii)$. First of all, we note that it follows from $(\aM)$ that $-C^{-1}\leq d\tilde{t}\cdot \altg^{-1}d\tilde{t}\leq - C$ for some $C>0$. We fix a cutoff function $\chi\in \coinf(\rr^{1+d})$ with $0\leq \chi\leq 1$, $\chi= 1$ near $0$ and set  $\chi_{R}(y)= \chi(R^{-1}y)$, $\hat{t}_{R}=\chi_{R} \tilde{t}+ (1- \chi_{R})t$.  We have:
\[
d\hat{t}_{R}= \chi_{R} d\tilde{t}+ (1- \chi_{R})dt+ (\tilde{t}- t)d\chi_{R}.
\]
The covector $\alpha_{R}= \chi_{R} d\tilde{t}+ (1- \chi_{R})dt$ is a convex combination of future directed timelike covectors, which using $(\aM) \ i)$ implies that there exists $C>0$ such that $-C^{-1}\leq \alpha_{R} \cdot \altg^{-1}\alpha_{R}\leq -C$, uniformly for $R\geq 1$. The error term $(\tilde{t}- t)d\chi_{R}$ is of norm $O(R^{-\epsilon})$, which shows that $\hat{t}_{R}$ is a  time function for $R$ large enough.  Let us fix such an $R$ and denote $\hat{t}_{R}$ by 
$\hat{t}$.  Clearly $\hat{t}-t\in \coinf(\rr^{1+d})$. It remains to check that $\hat{t}$ is a Cauchy time function. First using  that $\hat{t}=t+ \coinf(\rr^{1+d})$  we obtain that
\beq\label{debiloff}
\lim_{T\to +\infty}\sup_{\Sigma_{-T}}\hat{t}= -\infty, \ \ \lim_{T\to +\infty}\inf_{\Sigma_{T}}\hat{t}= +\infty.
\eeq
Let now $\gamma$ be an inextendible future directed continuous causal curve and $s\in \rr$. Since $\hat{t}$ is a time function, $\gamma$ intersects $\hat{t}^{-1}(s)$ at most once.  By global hyperbolicity, $\gamma$ intersects the Cauchy hypersurfaces $\Sigma_{\pm T}$  for  $T$ large enough. By \eqref{debiloff} this implies choosing $T$ very large that $\gamma$ intersects $\hat{t}^{-1}(s^{\pm})$ for some $s_{-}<s<s_{+}$ hence  also $\hat{t}^{-1}(s)$. Therefore $\hat{t}^{-1}(s)$ is a Cauchy hypersurface for each $s$ and $\hat{t}$ is a Cauchy time function. \qeds

\subsection{Reduction to the model case}\label{s11.1} We now perform the reduction to the model case following the method explained in \cite{inout}, here however we need to take into account the space-time decay of $\altg$ and $\altV$.

After possibly redefining $\tilde{t}$ by adding a constant, we can assume that $\Sigma\defeq\tilde{t}^{-1}(\{0\})= \{0\}\times \rr^{d}$, so that $\Sigma$ is a Cauchy hypersurface both for $\altg$ and  $\eta$.

We set $v= \dfrac{\altg^{-1}d\tilde{t}}{d\tilde{t}\cdot \altg^{-1} d\tilde{t}}$, so that $v= \p_{t}$ outside a compact set. If $\phi_{t}$ is the flow of $v$, we set:
\[
\chi: \rr\times \Sigma\in (t, \rx)\mapsto \phi_{t}(0, \rx)\in \rr^{1+d},
\]
so that $\tilde{t}(\chi(t, \rx))=t$. Thanks to the space-time decay properties $(\aM)$, the diffeomorphism $\chi$ has the following proprieties.

\begin{lemma}\label{l11.1}Assume $(\aM)$. Then
\[
\chi^{*}\altg= -  \altch^{2}(t, \rx)dt^{2}+ \hat  \alth(t, \rx)d\rx^{2}, \ \ \chi^{*}\altV=  \altVh, 
\]
where:
\[
\hat \alth, \hat \alth^{-1}, \altch, \altch^{-1}, \altVh\in S^{0}_{\std}(\rr^{1+d}).
\]
Moreover there exist  diffeomorphisms $\ry_{\outin}$ of $\Sigma$ with 
\[
\ry_{\outin}(\rx)- \rx\in S^{1-\delta}_{{\rm sd}}(\rr^{d}),
\]
such that if
\[
\hat{\alth}_{\outin}\defeq  \ry_{\outin}^{*}\altdelta,
\]
where $\altdelta$ is the flat Riemannian metric on $\rr^{d}$, we have:
\[
 \hat \alth- \hat \alth_{\outin}, \altch-1, \altVh- \altm^{2} \in S^{-\delta}_{\std}(\rr^{\pm}\times \rr^{d}).
\]
\end{lemma}
\proof  
We have $v= \p_{t}+ S^{-\delta}_{\std}$, which also implies that
\begin{equation}
\label{e11.1a}
\langle \phi_{s}(0, \rx)\rangle \geq C (\langle s\rangle+ \langle \rx\rangle),  \ \ C>0.
\end{equation}
Setting $w\defeq  \pi_{\ry}v$, we have $\pi_{\ry}\chi(t, \rx)= \pi_{\ry}\rx+ \int_{0}^{t}w(\phi_{s}(\rx))ds$. Using that $w\in S^{-\delta}_{\std}(\rr^{1+d})$, we obtain that
\[
\ry_{\outin}(\rx)\defeq  \lim_{t\to \pm\infty}\pi_{\ry}\chi(t, \rx)
\]
exist and:
\begin{equation}
\label{e11.2}
\pi_{\ry}\chi(t, \rx)-\ry_{\outin}(\rx)\in S^{1-\delta}_{\std}(\rr^{\pm}\times \rr^{d}), \ \ \ry_{\outin}(\rx)- \rx\in S^{1-\delta}_{{\rm sd}}(\rr^{d}).
\end{equation}
Since $\tilde{t}=t$ outside a compact set, we also have $\chi(t, \rx)= (t, \pi_{\ry}\chi(t, \rx))$ for $|t|+ |\rx|\geq C$, hence
\[
D\chi(t, \rx)= \mat{1}{0}{0}{D\ry_{\outin}(\rx)}+ S^{-\delta}_{\std}(\rr^{\pm}\times \rr^{d}).
\]
This estimate and \eqref{e11.1a} imply the assertion. \qeds

If $P\in {\rm Diff}(\rr^{1+d})$ we denote by $\chi^{*}P$ the pullback of $P$ by $\chi$ defined by  $(\chi^{*}P)u\circ \chi=(Pu)\circ \chi$.

We set $\hat{P}= \chi^{*}P$ and $\tilde{P}= \hat{c}^{1-n/2}\hat{P}\hat{c}^{1+n/2}$. By a direct computation, one finds
\beq\label{e11.333}
\tilde{P}= \pe_{t}^{2}+ r(t, \rx)\pe_{t}+ a(t, \rx, \pe_{\rx}), 
\eeq
where by Lemma \ref{l11.1}, $r$, $a$ satisfy $(\Hstd)$  for $\delta>1$, with 
\beq\label{e11.3b}
a_{\outin}(\rx, \pe_{\rx})\defeq -\Delta_{\hat{\alth}_{\outin}}+ \altm^{2}.
\eeq
Thus, one is reduced to the setting from previous sections, with the mere notational difference that the model Klein-Gordon operator from Sect. \ref{sec2} is now denoted $\tilde P$.

Note that 
\[
a_{\outin}(\rx, \pe_{\rx})= \chi_{\outin}^{*}(-\Delta_{\rx}+ \altm^{2}), \hbox{ where } \chi_{\outin}(t, \rx)\defeq (t, \ry_{\outin}(\rx)).
\]

There are several inconveniences related to the possibility that $a_\out\neq a_\inn$. It turns out, however, that they can be circumvented by considering the dynamics associated to the free Laplace operator $-\Delta_\rx+\altm^2$ instead of the asymptotic dynamics $\cU_\outin(t,s)$ associated to $a_{\outin}(\rx, \pe_{\rx})$.
 
\subsection{Wave operators}\label{wavu}
Let  $\Sigma_{s}\defeq \tilde{t}^{-1}(\{s\})$ for $\sobo\in \rr$. Using  the diffeomorphism $\chi$  we  identify $\Sigma_{s}$ with $\rr^{d}$  to define the Sobolev spaces $H^{\sobo}(\Sigma_{s})$. We introduce the energy spaces:
\[
\cE^{\sobo}(s)\defeq  H^{1+\sobo}(\Sigma_{s})\oplus H^{\sobo}(\Sigma_{s}), \ \sobo\in \rr.
\]
Of course,  under $\chi$ all spaces $\cE^{\sobo}(s)$ equal $\cE^{\sobo}(\rr^{d})$ with uniformly equivalent norms. 
We denote by $\cU(t,s): \cE^{\sobo}(s)\to \cE^{\sobo}(t)$ the Cauchy evolution associated to $P$. 

The approximate diagonalization of $\cU(t,s)$ can be performed after reduction to the model Klein-Gordon operator $\tilde P$, thus one needs to take into account the diffeomorphism $\chi$ and the conformal transformation involving appropriate powers of $\altch$. This is done in the lemma below, which is proved by straightforward computations detailed in \cite{inout}.

\begin{lemma} Let $\chi_t(\rx)=\ry(t,0,\rx)$. Let $Z(t): \cE^{\sobo}(t)\to \cH^{\sobo+\12}$ be defined by
\[
Z(t)\defeq (\chi_{t}^{*})^{-1} \altch^{n/2-1}\mat{1}{0}{-\i(n/2-1)\p_{t}\ln (\altch)}{1} \circ T(t),
\]
and similarly let us define $Z_{\outin}\defeq  (\chi_{\outin}^{*})^{-1}T_{\outin}$. Then
\begin{equation}
\label{e11.10}
\cU(t,s)= Z(t)\cU^{\adg}(t,s)Z^{-1}(s),
\end{equation}
where $\cU^{\adg}(t,s)$ is the approximately diagonalized Cauchy evolution of $\tilde P$ defined in \eqref{intolo}. Furthermore,
\beq\label{wlkj}
\lim_{t\to \pm \infty}Z(t)^{-1}Z_{\outin}-\one=0 \hbox{ in }B(\cH^{\sobo}).
\eeq
\end{lemma}

 We set $P_{\free}\defeq \pe_{t}^{2}- \Delta_{\rx}+ \altm^{2}$. Denoting by $\cU_{\free}(t,s)$ the usual Cauchy evolution for $P_{\free}$, we have by \eqref{e11.3b}:
\[
\cU_{\free}(t,s)= Z_{\outin}\cU^{\adg}_{\outin}(t,s)Z_{\outin}^{-1}.
\]
\begin{proposition}\label{propowavo}
 The limits
 \beq\label{e11.3bb}
W_{\outin}\defeq\lim_{t\to \pm\infty}\cU(0, t)\cU_{\free}(t, 0)
\eeq
exist in $B(\cE^{\sobo}(0))$ with inverses 
\beq\label{e11.3c}
W_{\outin}^{-1}=W_{\outin}^{\dag}= \lim_{t\to \pm\infty}\cU_{\free}(0, t)\cU(t, 0).
\eeq
Moreover one has:
\beq\label{e11.4c}
W_{\outin}= Z(0)W^{\adg}_{\outin}Z_{\outin}^{-1},
\eeq
where we recall that $W^{\adg}_{\outin}= \lim_{t\to\pm\infty}\cU^{\adg}(0,t)\cU^{\adg}_{\outin}(t,0)$.
\end{proposition}
\proof   The existence of the limits \eqref{e11.3bb}, \eqref{e11.3c} follows from the Cook argument, using the short range condition $\delta>1$. The  identity \eqref{e11.4c} follows from \eqref{wlkj}. \qed



\subsection{Fredholm problems and Feynman pseudo-inverse}\label{ss:fredf}
Following the notation in \eqref{e11.333}, the objects introduced in Sect. \ref{sec:abstract} will be denoted with tildes, like $\tilde{\cX}^{\sobo},\tilde{\cY}^{\sobo}$, etc. 
We define the spaces  
\beq
\bea
\cY^{\sobo}&\defeq (\chi^{-1})^{*}\tilde{\cY}^{\sobo}=(\chi^{-1})^{*}\big( \langle t\rangle^{-\gamma}L^{2}(\rr; H^{\sobo})\big),\\ 
\cX^{\sobo}&\defeq (\chi^{-1})^{*}\tilde{\cX}^{\sobo}=(\chi^{-1})^{*}\big\{ \tilde u \in \big(C^{1}(\rr; H^{\sobo+1})\cap C^{0}(\rr; H^{\sobo})\big): \ \tilde{P}\tilde{u} \in \tilde\cY^m\big\}.  
\eea
\eeq
In particular $\cX^{\sobo}$ is the space of $u\in \cD'(\rr^{1+d})$ such that $u\circ \chi\in C^{1}(\rr; H^{\sobo+1})\cap C^{0}(\rr; H^{\sobo})$ and $Pu\in \cY^{\sobo}$. We equip $\cY^{m}$ and $\cX^{m}$ with the norms obtained from $\tilde{\cY}^{m}$ and  $\tilde{\cX}^{m}$.

We now introduce the projections 
\[
c^{\pm, \vac}_{\free}=\12\begin{pmatrix}\one & \pm \sqrt{-\Delta_\rx+\altm^2} \\ \pm \sqrt{-\Delta_\rx+\altm^2} & \one\end{pmatrix}.
\] 
An easy computation shows that
\beq\label{e20.0}
c^{\pm, \vac}_{\free}= Z_{\out}\pi^{\pm}Z_{\out}^{-1}=Z_{\inn}\pi^{\pm}Z_{\inn}^{-1}.
\eeq

\begin{definition}\label{def:final}
 We set $\varo_{\outin}\defeq \slim_{t\to \pm \infty}\cU_{\free}(0,t)\varo_{t}$ and
 \[
 \bea
 \varo_{\F}\defeq c^{+,\vac}_{\free}\varo_{\out}+ c^{-, \vac}_{\free}\varo_{\inn}, \ \ W_{\F}^{\dag}\defeq c^{+,\vac}_{\free}W^{\dag}_{\out}+ c^{-, \vac}_{\free}W^{\dag}_{\inn},\\[2mm]
 \varo_{\bar{\F}}\defeq c^{-,\vac}_{\free}\varo_{\out}+ c^{+, \vac}_{\free}\varo_{\inn}, \ \ W_{\bar \F}^{\dag}\defeq c^{-,\vac}_{\free}W^{\dag}_{\out}+ c^{+, \vac}_{\free}W^{\dag}_{\inn},
 \eea
 \]
 and for $I\in \{\inn,\out,\F, \aF\}$:
 \[
 \cX^{\sobo}_{I}\defeq\{u\in \cX^{\sobo}: \ \varo_{I^{\rm c}}u=0\}.
 \]
 \end{definition}

\begin{theorem} Assume $(\aM)$ and let $P$, $\cX^\sobo_I$ be as defined in \eqref{e11.1} and Def. \ref{def:final} for $\sobo\in\rr$ and $I\in \{\inn,\out,\F, \aF\}$. Then:
\ben
 \item  $P: \cX^{\sobo}_{I}\to \cY^{\sobo}$ is Fredholm of index $\ind W_{I^{\rm c}}^{\dag}$, and invertible with inverse $G_{\pm}$ if $I= \outin$. Furthermore, $\Ker P|_{\cX^{\sobo}_{I}}\subset \cf(M)$ and the index does not depend on the Sobolev order $m$. 
 \item Let
 \[
 G_{\F}\defeq (\chi^{-1})^{*}(\altch^{1+n/2}\tilde{G}_{\F}\altch^{1-n/2}),
 \]
 where   $\tilde{G}_{\F}$ is the operator defined in \eqref{e100.20} and $\altch$, $\chi$ are defined in Subsect. \ref{s11.1}. Then:
\[
\begin{array}{rl}
i)&G_{\F}\in B(\cY^{\sobo}, \cX^{\sobo}_{\F}), \  PG_{\F}= \one_{\cY^{\sobo}}+ K_{\cY^{\sobo}}, \hbox{where }K_{\cY^{\sobo}}\hbox{ is compact on }\cY^{\sobo},\\[2mm]
ii)&G_{\F}P= \one_{\cX^{\sobo}_{\F}}+ K_{\cX^{\sobo}_{\F}}, \hbox{where }K_{\cX^{\sobo}_{\F}}\hbox{ is compact on }\cX^{\sobo}_{\F},\\[2mm]
iii)&\i^{-1}(G_{\F}- G_{\F}^{*})\geq 0\hbox{ on }\cY^{\sobo}, \hbox{ for }\sobo\geq 0,\\[2mm]
iv)&P G_{\F}-\one, G_{\F}P-\one \hbox{ are smoothing operators},\\[2mm]
v)& \WF(G_{\F})'= (\diag_{T^*M})\cup\textstyle\bigcup_{t\leq 0}(\Phi_t(\diag_{T^*M})\cap \pi^{-1}\cN).
\end{array}
\] 
In particular, $G_{\F}$ is a Feynman parametrix of $P$ in the sense of Def. \ref{def:Fp}.
 \een
 \end{theorem}
\proof  The maps \[
\begin{array}{l}
\cY^{\sobo}\ni f  \mapsto\tilde{f} \defeq \hat{c}^{1-n/2}f\circ \chi\in \tilde{\cY}^{\sobo},\\[2mm]
\cX^{\sobo}\ni u  \mapsto\tilde{u} \defeq \hat{c}^{1+n/2}u\circ \chi\in \tilde{\cY}^{\sobo},\\[2mm]
\end{array}
\]
are boundedly invertible and furthermore, $Pu=f$ iff $\tilde{P}\tilde{u}= \tilde{f}$. Moreover, by a direct computation we obtain that $Z^{-1}\varo u= T^{-1}\tilde{\varo}\tilde{u}$ and hence $u\in \cX_{I}^{\sobo}$ iff $\tilde{u}\in \tilde{\cX}_{I}^{\sobo}$. The theorem follows hence from Thm. \ref{teuheuteuheu}  provided we  check that 
\beq\label{finito}
\ind W_{I}^{\dag}= \ind W^{\adg\dag}_{I}.
\eeq
This is obvious for $I= \outin$ since  the  operators are then bijective. Let us check \eqref{finito} for $I=\aF$ for example. We denote   by $Z_{\free}$ the analog of $Z_{\outin}$ with $\epsilon_{\outin}$ replaced by $\epsilon_{\free}= (-\Delta_{\rx}+\altm^{2})^{\12}$ and $\chi_{\outin}$ replaced by $\one$.  Using \eqref{e11.4c} and \eqref{e20.0} we obtain that
\[
Z_{\free}^{-1}W_{\aF}^{\dag}= (Z_{\free}^{-1}Z_{\out}\pi^{-}W^{\adg\dag}_{\out}+ Z_{\free}^{-1}Z_{\inn}\pi^{+}W^{\adg\dag}_{\inn}) Z(0)^{-1}=  S\circ W_{\aF}^{\adg\dag}\circ Z(0)^{-1},
\] 
for $S= Z_{\free}^{-1}Z_{\out}\pi^{-}+ Z_{\free}^{-1}Z_{\inn}\pi^{+}$. But since 
$c^{\pm, \vac}_{\free}= Z_{\free}\pi^{\pm}Z_{\free}^{-1}$, $Z_{\free}^{-1}Z_{\outin}$ commutes with $\pi^{+}$ and $\pi^{-}$, using again \eqref{e20.0}. Therefore $S= \pi^{-}Z_{\free}^{-1}Z_{\out}\pi^{-}+ \pi^{+}Z_{\free}^{-1}Z_{\inn}\pi^{+}$ is invertible and hence $\ind W^{\dag}_{\aF}= \ind W^{\adg\dag}_{\aF}$. \qeds

\appendix
\section{}\init\label{secapp1}
\subsection{Proof of Prop. \ref{l5.1}}\label{ssecap1}
To prove Prop. \ref{l5.1} we first need an auxiliary lemma about parameter-dependent pseudo\-differential calculus.

 We start by introducing parameter dependent versions of the $\Psi^{m, 0}_{\std}(\rr; \rr^{d})$ calculus.  Namely, we  define $\widetilde{S}_{\std}^{m, \delta}(\rr; T^{*}\rr^{d})$ to be the space of functions $c(t, \rx, \spexi, \lambda)$ such that:
\[
\p_{t}^{n}\p_{\lambda}^{p}\p_{\rx}^{\alpha}\p_{\spexi}^{\beta}c(t, \rx, \spexi, \lambda)\in O\big((\langle x\rangle + \langle t \rangle)^{\delta-n-|\alpha|}(\langle \spexi\rangle + \langle \lambda\rangle)^{m-|\beta|- p}\big), \ \alpha, \beta\in \nn^{d}, \ p, n\in \nn.
\]
The typical element of $\widetilde{S}_{\std}^{m,0}(\rr; T^{*}\rr^{d})$ is $c(t, \rx, k, \lambda)= a(t, \rx, k)+ \lambda^{m}$ for $a\in S_{\std}^{m, 0}(\rr; T^{*}\rr^{d})$. 

We denote by $\widetilde{S}_{\std, {\rm ph}}^{m, \delta}(\rr; T^{*}\rr^{d})$ the subspace of symbols wich are polyhomogeneous in $(k, \lambda)$.

Furthermore, we define $\widetilde{\cW}^{-\infty}_{\std}(\rr; \rr^{d})$ as the set of  maps $\rr\ni \lambda\mapsto a( t, \lambda)\in \cW^{-\infty}_{\std}(\rr; \rr^{d})$ such that:
\[
\|(D_{\rx}^{2}+\rx^{2}+\lambda^{2}+1)^{m}\p_{t}^{n}\p_{\lambda}^{p}a(t, \lambda)(D_{\rx}^{2}+\rx^{2}+\lambda^{2}+1)^{m}\|_{B(L^{2}(\rr^{d}))}\in O(\langle t\rangle^{-m}\langle \lambda\rangle^{-m}), 
\]
for all $m,n,p\in \nn$. We set
\[
\widetilde{\Psi}^{m, \delta}_{\std}(\rr; \rr^{d})= \Op\big(\widetilde{S}_{\std, {\rm ph}}^{m, \delta}(\rr; T^{*}\rr^{d})\big)+ \widetilde{\cW}^{-\infty}_{\std}(\rr; \rr^{d}).
\]

\begin{lemma}\label{lemomo}
 Suppose that $a(t)\in\Psi^{2, 0}_{\std}(\rr; \rr^{d})$ and that $a(t)$ is elliptic, selfadjoint on $L^{2}(\rr^d)$ and satisfies $a(t)\geq c_{0}\one$,  $c_{0}>0$. Then $(a(t)+ \lambda^{2})^{-1}\in\widetilde{\Psi}^{-2, 0}_{\std}(\rr; \rr^{d})$.
 \end{lemma}

\proof  Let us consider the operator  $A(t)= a(t)+ D_{l}^{2}$ acting on $L^{2}(\rr^d\times\rr)$, where $l$ is the dual variable to $\lambda$. It is selfadjoint on $H^{2}(\rr^d\times \rr)$ and satisfies $A(t)\geq c_{0}\one$. The idea of the proof is to construct a  time-dependent pseudodifferential calculus on $\rr^{d}_{\rx}\times \rr_{l}$ in which $A(t)$ is elliptic, and such that Seeley's theorem holds true. 

{\it Step 1}. In step 1 we construct  a convenient pseudodifferential calculus acting on $\rr^{d+1}= \rr^{d}_{\rx}\times \rr_{l}$.
Specifically, the symbol classes $S^{m}_{\std}(\rr; T^{*}\rr^{d+1})$ are defined by the conditions
\[
\p_{t}^{n}\p_{\rx}^{\alpha}\p_{\spexi}^{\beta}\p_{l}^{p}\p_{\lambda}^{m}a(t, \rx, \spexi, l, \lambda)\in O\big((\langle \rx\rangle+ \langle t\rangle)^{-n-|\alpha|}(\langle \spexi\rangle+ \langle \lambda\rangle)^{m-|\beta|-m}\big), 
\]
for all $\alpha, \beta\in \nn^{d}, \ m,n, p\in \nn$. We denote by $S^{m}_{\std, {\rm ph}}(\rr; T^{*}\rr^{d+1})$ the subspace of symbols which are polyhomogeneous in $(k, \lambda)$.

The class  of $t-$dependent pseudodifferential operators $\widetilde{\Psi}^{m}_{\std}(\rr; \rr^{d+1})$
consists by definition of sums of   $\widetilde{\rm Op}(S^{m}_{\std, {\rm ph}}(\rr; T^{*}\rr^{d+1}))$, where $\widetilde{\rm Op}$ denotes the Weyl quantization in $(x, l)$, and of elements of the ideal 
 $\cW_{\std}^{-\infty}(\rr; \rr^{d+1})$  of operator-valued functions $\rr\ni t\mapsto a(t)$ such that
\beq\label{difito}
\|(D_{l}^{2}+D_{\rx}^{2}+  \langle\rx\rangle^{2})^{n}\p_{t}^{m}{\rm ad}_{l}^{p}a(t)(D_{l}^{2}+ D_{\rx}^{2}+ \langle\rx\rangle^{2})^{n}\|_{B(L^{2}(\rr^{d}_{\rx}\times \rr_{l}))}\in O(\langle t\rangle^{-n}),
\eeq
 for all $m,n,p\in \nn$, where we use the notation ${\rm ad}_{A}B= [A, B]$. 

Let us check that Seeley's theorem is still valid for this class of operators, by verifying once again the abstract conditions in \cite{alnv1}. As in the proof of Thm. \ref{seeley-std} we use the notations in \cite[Subsect. 5.3]{bounded}.
 We take as Hilbert space $\cH= L^{2}(\rr_{t}\times \rr^{d+1})$, as injective operator in $\cW_{\std}^{-\infty}(\rr; \rr^{d+1})$ the operator $\e^{- D_{\rx}^{2}+ D_{l}^{2}+ \rx^{2}+ t^{2}+1}$.   The fact that $a\in S^{m}_{\std, {\rm ph}}(\rr; T^{*}\rr^{d+1})$ and $\widetilde{\rm Op}(a)\in \cW_{\std}^{-\infty}(\rr; \rr^{d+1})$ implies $a\in S^{-\infty, 0}_{\std, {\rm ph}}(\rr; T^{*}\rr^{d+1})$ is easy to check.
 
The proof of the spectral invariance of $\cW_{\std}^{-\infty}(\rr; \rr^{d+1})$ is done as in Thm. \ref{seeley-std}, using that the operator ${\rm ad}_{l}$ satisfies Leibniz rule and the identity ${\rm ad}_{l}(\one - R)^{-1}= (\one - R)^{-1}{\rm ad}_{l}R(\one - R)^{-1}$.

   {\it Step 2}.  In step 2 we describe the 
  relationship between  the classes $\Psi_{\std}^{m}(\rr; \rr^{1+d})$ and $\widetilde{\Psi}_{\std}^{m}(\rr; \rr^{d})$.
   Denoting by $\mathcal{F}$ the Fourier transform in $l$ and by $T_{l}$ the group of translations in $l$, we see that
\[
\begin{array}{rl}
&c\in S^{m}_{\std, {\rm ph}}(\rr; T^{*}\rr^{1+d}),\ [T_{l}, \widetilde{\rm Op}(c)]=0\\[2mm]
\Leftrightarrow &
\mathcal{F} \Op(c)\mathcal{F}^{-1}= \int^{\oplus}_{\rr}\Op(c(\lambda))d\lambda, \hbox{ for }c(\lambda)\in \widetilde{S}^{m}_{\std, {\rm ph}}(\rr; T^{*}\rr^{d}).
\end{array}
\]
Let now $w\in \cW_{\std}^{-\infty}(\rr; \rr^{d+1})$ with $[w, T_{l}]=0$.  We have:
 \beq\label{four}
 \mathcal{F}w\mathcal{F}^{-1}= \int^{\oplus}_{\rr} w(t,\lambda)d\lambda.
 \eeq
 From \eqref{difito} we obtain that:
 \[
 \bea
&\int_{\rr}\langle \lambda\rangle^{n}\|(D_{\rx}^{2}+ \langle\rx\rangle^{2})^{n}\p_{t}^{m} \p^{p}_{\lambda}w(t,\lambda)u(\lambda)\|^{2}_{L^{2}(\rr^{d})}d\lambda\\[2mm]
&\leq C_{n, p}\langle t\rangle^{-n}\int_{\rr}\| (D_{\rx}^{2}+ \langle\rx\rangle^{2})^{-n}u(\lambda)\|^{2}_{L^{2}(\rr^{d})}d\lambda,\ \ \forall\, m,n, p\in \nn,
\eea
 \]
 or equivalently
 \[
 \int^{\oplus}_{\rr}\langle \lambda\rangle^{n}(D_{\rx}^{2}+ \langle \rx\rangle^{2})^{n/2}\p_{t}^{m}\p_{\lambda}^{p}w(\lambda)(D_{\rx}^{2}+ \langle\rx\rangle^{2})^{n/2}d\lambda
 \]
 is of norm $O(\langle t\rangle^{-n})$ in $B(L^{2}(\rr^{d+1}))$.
 By Sobolev's embedding theorem this  implies that
 \[
 \|(D_{\rx}^{2}+ \langle \rx\rangle^{2})^{n/2}\p_{t}^{m}\p_{\lambda}^{p}w(\lambda)(D_{\rx}^{2}+ \langle\rx\rangle^{2})^{n/2}\|_{B(L^{2}(\rr^{d}))}\in O(\langle t\rangle^{n}\langle \lambda\rangle^{-n})\ \forall m,n, p\in \nn,
 \]
  hence $w(t,\lambda)\in \widetilde{\cW}_{\std}^{-\infty}(\rr; \rr^{d})$.
 Conversely, if $w(t,\lambda)\in  \widetilde{\cW}^{-\infty}_{\std}(\rr; \rr^{d})$  it is immediate that  $w$ defined by \eqref{four}
 belongs to $\cW_{\std}^{-\infty}(\rr; \rr^{d+1})$. Hence we have shown
 \begin{equation}
 \label{e.deco2}
 \bea
 & w\in \cW_{\std}^{-\infty}(\rr: \rr^{d+1}), \ \ [w, T_{l}]=0\\[2mm]
 \Leftrightarrow& \ \mathcal{F}w\mathcal{F}^{-1}= \int^{\oplus}_{\rr}w(\lambda)d\lambda, \hbox{ for }  w(\lambda)\in \widetilde{\cW}_{\std}^{-\infty}(\rr; \rr^{d}). 
 \eea
  \end{equation}

{\it Step 3}. We can now conclude the proof of the lemma. 
   By assumption  $A(t)= a(t)+ D_{l}^{2}\in\Psi^{2}_{\std}(\rr; \rr^{d+1})$ and is uniformly elliptic  in that class, hence by Seeley's theorem  $A(t)^{-1}\in\Psi^{-2}_{\std}(\rr; \rr^{d+1})$.  We have
   \[
\mathcal{F}A(t)^{-1}\mathcal{F}^{-1}= \int^{\oplus}_{\rr}(a(t)+ \lambda^{2})^{-1}d\lambda,
\]
where by Step 2 $(a(t)+ \lambda^{2})^{-1}\in \widetilde{\Psi}_{\std}^{-2}(\rr; \rr^{d})$. This completes the proof of the lemma. \qed

\noindent {\bf Proof of Prop.  \ref{l5.1}} In view of the identity
\[
a_{1}^{1+ \alpha}- a_{2}^{1+ \alpha}= (a_{1}- a_{2})a_{1}^{\alpha}+ a_{2}(a_{1}^{\alpha}- a_{2}^{\alpha}),
\]
 we see that it suffices to prove the proposition for $0<\alpha<1$. We  will use the following formula, valid for example  if $a$ is a selfadjoint operator on a Hilbert space $\cH$  with $a\geq c\one$, $c>0$:
\beq\label{powers}
a^{\alpha}= C_{\alpha}\int_{0}^{+\infty}(a+ s)^{-1}s^{\alpha}ds= C_{\alpha}\int_{\rr}(a+ \lambda^{2})^{-1}\lambda^{2\alpha+1}d\lambda, \ \alpha\in \rr,
\eeq
where the integrals are norm convergent in say, $B(\Dom a^{m}, \cH)$ for $m$ large enough.

We have for $r(t)= a_{1}(t)- a_{2}(t)$:
\[
\bea
(a_{1}(t)+ \lambda^{2})^{-1}&= ( a_{2}(t)+ \lambda^{2})^{-1}(\one + r(t)(a_{1}(t)+ \lambda^{2})^{-1})\\
&= ( a_{2}(t)+ \lambda^{2})^{-1}+  ( a_{2}(t)+ \lambda^{2})^{-2}(a_{2}(t)+ \lambda^{2})r(t)(a_{1}(t)+ \lambda^{2})^{-1}\\
&= (a_{2}(t)+ \lambda^{2})^{-1}+ (a_{2}(t)+ \lambda^{2})^{-2}a_{2}(t)c_{1}(t, \lambda)\\
&= (a_{2}(t)+ \lambda^{2})^{-1}+ a_{2}(t)c_{2}(t, \lambda), 
\eea
\]
where using Lemma \ref{lemomo}, $c_{1}(t, \lambda)\in \widetilde{\Psi}^{0, - \delta}(\rr; \rr^{d}))$ and $c_{2}(t, \lambda)\in  \widetilde{\Psi}^{-4, - \delta}(\rr; \rr^{d})$. From \eqref{powers} we obtain that:
\beq\label{potopoto}
a_{1}^{\alpha}(t)- a_{2}^{\alpha}(t)= C_{\alpha}a_{2}(t)\int_{\rr}c_{2}(t, \lambda)\lambda^{2\alpha+1}d\lambda.
\eeq
We now write $c_{2}(t, \lambda)$ as $\Op(d_{2}(t, \lambda))+ w_{2}(t, \lambda)$, for $d_{2}\in \widetilde{S}_{\std, \rm ph}^{-4,-\delta}(\rr; T^{*}\r^{d})$ and $w_{2}(t, \lambda)\in \widetilde{\cW}^{-\infty}(\rr; \rr^{d})$.
Using that
\[
\int_{\rr}(\langle \xi\rangle + \langle \lambda \rangle)^{-4-k}\lambda^{2\alpha+1}d\lambda\sim \langle \xi\rangle^{2\alpha-2-k},
\]
we first obtain that
\[
\int_{\rr}d_{2}(t, \lambda)\lambda^{2\alpha+1}d\lambda\in S_{\std, \rm ph}^{2\alpha-2, -\delta}(\rr; \rr^{d}).
\]
Similarly we obtain that $\int_{\rr}w_{2}(t, \lambda)\lambda^{2\alpha+1}d\lambda\in \cW^{-\infty}_{\std}(\rr; \rr^{d})$.  Using \eqref{potopoto} this implies that $a_{1}^{\alpha}(t)- a_{2}^{\alpha}(t)\in\Psi^{2\alpha, - \delta}(\rr; \rr^{d})$, as claimed. \qed

\subsection{Proof of Lemma \ref{turlututi}}\label{apota}
\newcommand\epsi[1]{\langle \epsilon(#1)\rangle}
\def\tpx{(\langle \rx\rangle + \langle t \rangle)}
By interpolation, it suffices to prove the lemma for $m, k\in \nn$. Let us set
\[
\begin{array}{l}
T_{m,k}(t)= \epsi{0}^{m}\langle \rx\rangle^{k}\cU^{\adg}(0,t)\tpx^{-k}\epsi{0}^{-m},\\[2mm]
R_{m,k}(t,s)= \cU^{\adg}(t,s) \epsi{s}^{m}\langle \rx\rangle^{k}\cU^{\adg}(s,t)\tpx^{-k}\epsi{0}^{-m}.
\end{array}
\]
Using the  uniform ellipticity of $\epsilon(t)$  it suffices to prove that
\beq\label{e.turo-1}
\sup_{t\geq 0}\| T_{m,k}(t)\|_{B(\cH^{0})}<\infty.
\eeq We claim that
\begin{equation}
\label{e.turo1}
\sup_{0\leq s\leq t}\| R_{m, k}(t,s)\|_{B(\cH^{0})}<\infty,
\end{equation}
This of course implies   \eqref{e.turo-1}  by taking $s=0$ and using that $\cU^{\adg}(t,s)$ is uniformly bounded in $B(\cH^{0})$, see \cite{inout}. To prove \eqref{e.turo1} we compute
 \beq\label{e.turo0}
\bea
&\p_{s}R_{m,k}(t,s)\\
&= \cU^{\adg}(t,s)\left(\p_{s}\epsi{s}^{m}+ [H^{\adg}(s), -\i \epsi{s}^{m}] \right)\langle \rx\rangle^{k}\cU^{\adg}(s,t)\tpx^{-k}\epsi{0}^{-m}\\
&\phantom{=}+ \cU^{\adg}(t,s)\epsi{s}^{m}[H^{\adg}(s), -\i \langle \rx\rangle^{k}]\cU^{\adg}(s,t)\tpx^{-k}\epsi{0}^{m}.
\eea
\eeq
Recall that
\[
H^{\adg}(t)= \mat{\epsilon(t)}{0}{0}{-\epsilon(t)}+ \Psi_{\std}^{0, -1-\delta}(\rr; \rr^d)\otimes B(\cc^{2})
\]
by Prop. \ref{propoesti}. Hence:
\[
\left(\p_{s}\epsi{s}^{m}+ [H^{\adg}(s), -\i \epsi{s}^{m}]\right) \in \Psi^{m, -1- \delta}_{\std}(\rr; \rr^d)\otimes B(\cc^{2}),
\]
and we can write:
\begin{equation}
\label{e.turo2}
\left(\p_{s}\epsi{s}^{m}+ [H^{\adg}(s), -\i \epsi{s}^{m}]\right)= A_{m}(s)\epsi{s}^{m}\langle \rx\rangle^{-1},
\end{equation}
where 
\beq\label{e.turo3}
\|A_{m}(s)\|_{B(\cH^{0})}\in O(1),
\eeq
since  $\tpx^{-1- \delta}\leq \langle \rx\rangle^{-1}$. Similarly, we have 
\beq\label{e.turo4}
\epsi{s}^{m}[H^{\adg}(s), \i \langle \rx\rangle^{k}]=  C_{m, k}(s)\epsi{s}^{m}\langle \rx\rangle^{k-1}, 
\eeq
where
\begin{equation}
\label{e.turo5}
\|C_{m,k}(s)\|_{B(\cH^{0})}\in O(1).
\end{equation}
We also set
\[
B_{m,k}(t)= \tpx^{-k+1}\epsi{0}^{-m}\tpx^{k-1}\epsi{0}^{m},
\]
and we have by pseudodifferential calculus
\begin{equation}
\label{e.turo6}
\|B_{m,k}(t)\|_{B(\cH^{0})}\in O(1).
\end{equation}
Hence, we can rewrite \eqref{e.turo0} as
\beq\label{e.turo7}
\bea
&\p_{s}R_{m,k}(t,s)\\
&=\cU^{\adg}(t,s)D_{m,k}(s)\cU^{\adg}(s,t)\times R_{m, k-1}(t,s)\times B_{m,k}(t)\times \tpx^{-1},
\eea
\eeq
where
\beq\label{e.turo10}
D_{m,k}(s)= A_{m}(s)+ C_{m, k}(s), \ \ \|D_{m, k}(s)\|_{B(\cH^{0})}\in O(1).
\eeq
We can prove now \eqref{e.turo1} by induction for  $k$. First, note that by 1) of \cite[Prop. 5.6]{inout}, \eqref{e.turo1} holds for $k=0$. Assume that \eqref{e.turo1} holds for $k-1$. Integrating  \eqref{e.turo7} from $t$ to $s$ we obtain:
\[
\bea
\|R_{m,k}(t,s)- R_{m, k}(t,t)\|\leq \int_{0}^{t}\|R_{m, k-1}(t, \sigma)\|\langle t\rangle^{-1}dt\in O(1), \hbox{for }0\leq s\leq t 
\eea
\]
by the induction hypothesis. We conclude the proof of \eqref{e.turo1} using that
\[
\|R_{m, k}(t,t)\|=\| \epsi{t}^{m}\langle \rx\rangle^{k}\tpx^{-k}\epsi{0}^{m}\|\in O(1). 
\]
This completes the proof of the lemma. \qeds

\subsection*{Acknowledgments} It is a pleasure to thank Jan Derezi\'nski, Peter Hintz, Daniel Siemssen, Alexander Strohmaier, Andr\'as Vasy and Jochen Zahn for stimulating and helpful discussions. M.\,W. gratefully acknowledges the financial support of the National Science Center, Poland, under the grant UMO-2014/15/B/ST1/00126. The authors also wish to thank the Erwin Schr\"odinger Institute in Vienna for its hospitality during the program ``Modern theory of wave equations'', which greatly helped in the completion of this work.

\end{document}